\documentclass[a4paper,twoside]{article}
\usepackage[authoryear]{natbib}
\bibpunct{(}{)}{;}{a}{}{,}
\usepackage{graphicx}
\usepackage{subfigure}
\date{} 
\newcommand{\kms}{\mbox{km\,s$^{-1}$}}
\newcommand{\ergs}{\mbox{ergs\,s$^{-1}$}}
\title{\large\bf\flushleft 
Resolving the Structure at the Heart of BAL Quasars Through
Microlensing Induced Polarisation Variability
}
\author{\parbox{\textwidth}{\flushleft
\vspace{-0.5cm}
{\it Christopher A. Hales$^{A}$ and Geraint F. Lewis}\\
\vspace{0.4cm}
{\small Institute of Astronomy, School of Physics, University of Sydney, NSW 2006, Australia}\\
{\small $^{A}$chales@physics.usyd.edu.au}}}
\begin{document}
\twocolumn[
\begin{changemargin}{.8cm}{.5cm}
\begin{minipage}{.9\textwidth}
\vspace{-1cm}
\maketitle
%
%
\small{\bf Abstract:} While amongst the most luminous objects in the universe,
many  details  regarding  the inner structure of quasars  remain unknown.  One
such  area is  the mechanism  promoting  increased polarisation  in the  broad
absorption   line   troughs   of   certain   quasars.   This  study shows  how
microlensing  can be  used to  differentiate between  two popular  models that
explain  such  polarisation through  a  realistic  computational analysis.  By
examining a  statistical ensemble of correlation data  between two observables
(namely image brightness and polarisation of the flux coming from the quasar),
it was  found that through spectropolarimetric monitoring it would be possible
to  discern between a  model with  an external  scattering region  and a model
without one.


\medskip
\medskip
\end{minipage}
\end{changemargin}
]
\small

\section{Introduction}
Quasars are amongst the most  luminous objects in the universe, radiating most
of their energy from within a small continuum emitting region only $\sim1$ pc in
extent.   Such  a source  at  cosmological  distances  subtends the  order  of
microarcseconds,  well below  the highest  angular resolution  obtainable with
modern  telescopes. However,  gravitational  microlensing can  be employed  to
reveal  the structure  at  the heart  of  quasars, with  magnification due  to
individual  stars  revealing  information  about  the size  of  the  continuum
emitting  accretion  disk \citep{wam:4,wit,haw,gou}  and  broad emission  line
region   \citep{2002ApJ...577..615W,2002ApJ...576..640A,lew:2}.   

While  the microlensing  advances have  improved our  understanding  of quasar
structure, the  picture  is  far  from  complete, especially  with  regards to
areas such  as the  mechanism promoting jet  activity [which in  turn dictates
radio-loudness  or -softness, e.g.  \citet{kun,cat,fen}] and  its relation  to
prominent absorbing  regions \citep[e.g.][]{mur:1,lew:1}.  In  particular, the
exact mechanism which promotes  increased polarisation in the broad absorption
line (BAL) troughs of quasar spectra  is not well understood.  This paper will
extend  recent  studies \citep{lew:1,bel:2}  to  computationally simulate  how
gravitational microlensing can be used  to discern between the two main models
for polarisation  enhancement and hence probe  the scales of  structure in the
absorbing/scattering   regions.   Section  2   discusses  of  the  details  of
BAL  quasars, microlensing, and the  quadruply imaged  quasar  H1413+1143, the
system which is the ideal  observational candidate for this study.  Section  3
details the  approach to  the simulations, including  the construction  of the
source profiles  and polarisation maps,  with the results  and  conclusions of
this study presented in Sections 4 and 5 respectively.

\section{Background}

\subsection{BAL Quasars and Polarisation}\label{bals}
Approximately  10-20\%  of  optically  selected  quasars  have  been found  to
exhibit  broad   absorption  troughs  in  resonant  lines   blue-ward  of  the
corresponding   emission  lines   \citep{hew,rei},  exhibiting   bulk  outflow
velocities of  around 5000-30000 \kms\  \citep{tur:1}. These BAL  quasars were
thought  to  consist  of  only  radio-quiet or radio-intermediate sources, but
the discovery  of a radio-loud BAL  quasar  suggests   the  phenomenon  occurs
throughout  the quasar  population \citep{bec}.   Current theories  suggest an
orientation-based unification  scheme to explain  the occurrence of the  BAL in
quasars (e.g.  \citet{ant:1}), and of  interest to this study is the scattering
structure   which  increases  polarisation   within  BAL  troughs.

\begin{figure*}
\begin{center}
\includegraphics[angle=0,width=0.6\textwidth]{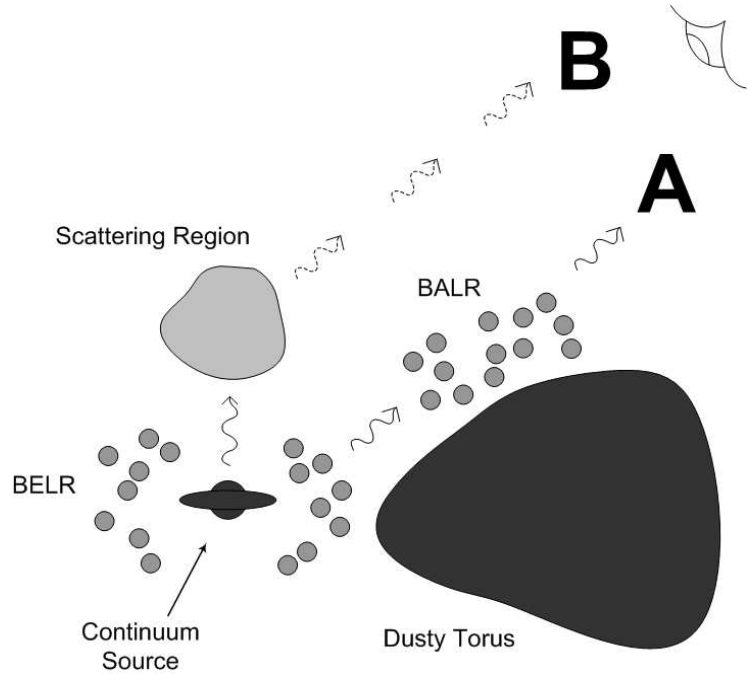}
\caption{The two potential models for the scattering and absorption structure
 (see explanation in Section \ref{bals}; also \citet{bel:2}).}\label{fig2}
\end{center}
\end{figure*}

Figure \ref{fig2} shows  the two competing models for  explaining the enhanced
polarisation in the broad absorption lines \citep{coh,ogl,schm}. For Model  A,
radiation from  the nuclear region only  travels along Path A  through the BAL
clouds.  This radiation  is then enhanced by resonant  scattering
into  the line  of sight,  increasing the  polarisation within  the absorption
troughs.  For Model B, radiation travels along Paths A and B. However, in this
case  Path A  does not  necessarily  introduce any  increase in  polarisation.
Radiation travelling  along Path B is  scattered into the line  of sight, most
likely from  a region of electrons and  dust \citep{ant:2,goo,gal,bra}.  Here,
the increased  polarisation in  the absorption troughs  is due to  the reduced
amount of  unpolarised flux  coming directly from  the continuum along  Path A
(the polarised  flux remains  relatively free of  absorption as it  avoids the
BAL clouds).

\subsection{Microlensing}
The details  of gravitational lensing can  be found in  recent review articles
\citep[e.g.][]{1998LRR.....1...12W}  and  only the  salient  features will  be
discussed here.  Gravitational  lensing occurs when light rays  from a distant
source  pass near  a massive  object  and suffer  achromatic deflection.   The
deflection angle $\alpha$  of such a light ray passing at  a distance $r$ from
an object of mass  $m$ is given by Equation \ref{eq1} -  i.e.  light rays will
follow null geodesics in the presence of massive objects.
\begin{equation}
\alpha = \frac{4Gm}{c^{2}r}\label{eq1}
\end{equation} 
The  measurement of this  deflection was  one of  the first  key observational
tests of Einstein's general relativity.

When considering the  deflections due to individual galaxies,  it is seen that
multiple  light paths  can connect  a source  with an  observer,  resulting in
multiple  imaging  on the  scale  of  arcseconds.   However, when  small-scale
granularity  in the  distribution of  galactic matter  (stars,  planets, black
holes) is  considered, it is seen  that these macrolensed  images are actually
composed of a myriad of unresolvable ($\sim10^{-6}$ arcsec) microimages due to
imaging  by stellar-mass  objects \citep{chan}.   While these  microimages are
unresolvable,  the  stellar mass  objects  can  introduce  magnification of  a
background  source,  and  the  motions   of  the  lensing  stars  can  produce
significant fluctuations into the  observed brightness of the macroimage.  The
characteristic scale length of microlensing is the Einstein radius (ER). For a
point  mass  lensing,  and  a  perfect  alignment  of source (s), lens (l) and
observer (o), the result  would  be a  circular  image  known  as  an Einstein
ring\footnote{Despite being first proposed by \citet{chw}} at the Einstein Radius
(ER). The  physical  projection  of this  radius  onto the  source is given by
Equation \ref{eq2}, where D is the angular diameter distance.
\begin{equation}
ER=\sqrt{\frac{4Gm}{c^{2}}\frac{D_{os}D_{ls}}{D_{ol}}}\label{eq2}
\end{equation}
The importance of this length scale is that objects smaller  than it  are much
more susceptible to large magnifications,  while objects larger than it suffer
less magnification \citep{wam:3,schn,wam:2}.

\subsection{H1413+1143}\label{h1413}
H1413+1143 (the Cloverleaf) consists of  4 images of a $z=2.55$ quasar
with  angular separations  of 0.77  to 1.36  arcsec \citep{mag,tur:3}.
The    lensing   galaxy    has   only    recently    been   identified
\citep{1998A&A...339L..65K,chant},      with      \citet{mag}      and
\citet{1990A&A...233L...5A}   identifying  two   prominent  absorption
systems at  $z=1.438$ and 1.661; for  the purposes of  this study, and
for consistency  with \citet{lew:1},  we adopt the  mean of  these two
values,  $z=1.55$, to represent  the redshift  of the  lensing galaxy.
One of  the images,  D, (see \citet{char}),  has been seen  to exhibit
variability           consistent           with           microlensing
\citep{1990A&A...233L...5A,kay,ost},   with  \citet{hut:1}  suggesting
that prominent  differences in the  absorption profiles of  the images
might be due  to selective microlensing of absorbing  clouds, with the
scale size  of these clouds  being smaller than  the continuum-forming
region.   It was  conceded, however,  that this  would require  a very
precise lensing configuration.   Furthermore, the polarisation for the
summed  images  has  been  seen  to  vary in  the  blue  wing  of  the
C$_{\textrm{IV}}$  $\lambda$1549   emission  line,  with  fluctuations
between   $\sim$10\%    \citep{lam}   and   $\sim$20\%   \citep{schm}.
Interestingly,  these studies also  found that  continuum polarisation
near  the  C$_{\textrm{IV}}$ feature  was  $\sim$2\%, indicating  some
polarisation  is  still occurring  away  from  the absorption  troughs
\citep{wan}.

There  seems  to  be growing  support  in  the  literature that  an  external,
asymmetric  scattering region  is responsible  for this  polarisation increase
+(Model  B   in   Section~\ref{bals}).   A recent  $HST$  study  of  H1413+1143
\citep{chae:2}  has indicated  that  the size  scale  for such  a region  lies
approximately between the  Einstein ring size (i.e.  Einstein  diameter in the
source plane) and $10^{18}L_{46}^{0.5}$ cm (where $L_{46}$ is the lensed quasar
luminosity  in  units  of $10^{46}$ \ergs\ ). The  minimum value comes from the
requirement  that  the scattering  region  should  not  be as  susceptible  to
microlensing as  the central nucleus, whereas  the maximum value  is simply an
estimate for the  size scale  of  the  broad  emission   line   region  (BELR)
\citep{mur:2,kas}.  If   the  scattering  region  were  any  larger  then  the
differences  seen  between  macrolensed  images  would  not   be   as   great.
Furthermore, recent data  from \it Chandra \rm \citep{char} appears to  support this
model.

By applying  more realistic models for  both the BAL region  of H1413+1143 and
the  distribution  of stars  in  the  lensing  galaxy, the  observed  spectral
variations   examined   by   \citet{hut:1}   could   be   readily  reproduced,
removing the  need for precise lensing configurations \citep{lew:1}. This idea
was  furthered by \citet{bel:2} who investigated  the role that Models A and B
(see Figure  \ref{fig2}) might play  in explaining the  increased polarisation
within   BAL  troughs.   However,   this  previous   study  only   tested  one
configuration  with a scattering region scale size of half an Einstein radius.
Hence,   how  secure  are   the  conclusions   drawn  by   \citet{chae:2}  and
\citet{char}, and could Model A  produce polarisation variations that could be
misinterpreted as Model  B? To this end, the remainder  of this paper examines
detailed  simulations  of  both  models  and the  predictions  they  make  for
polarisation variability.

\section{Method} 
\subsection{Ray Tracing}\label{raytrac}
When considering microlensing at cosmological scales, many stars influence the
path of  light through a galaxy,  and the single,  isolated lens approximation
which works well in Galactic  microlensing must be abandoned. Such a situation
is  analytically intractable and  numerical techniques  must be employed. This
study employs  the inverse/backwards ray-shooting  technique \citep{kay,wam:1}
which  involves `shooting' light  rays from  the observer  to the  lens plane,
calculating the deflection due to  individual stars in the lensing galaxy, and
then  collecting these  rays in  the  pixels of  the source  plane where  they
eventually hit, forming a magnification map.

\begin{figure*}
\begin{center}
\begin{minipage}[c]{0.25\textwidth}
\includegraphics[angle=0,width=\textwidth]{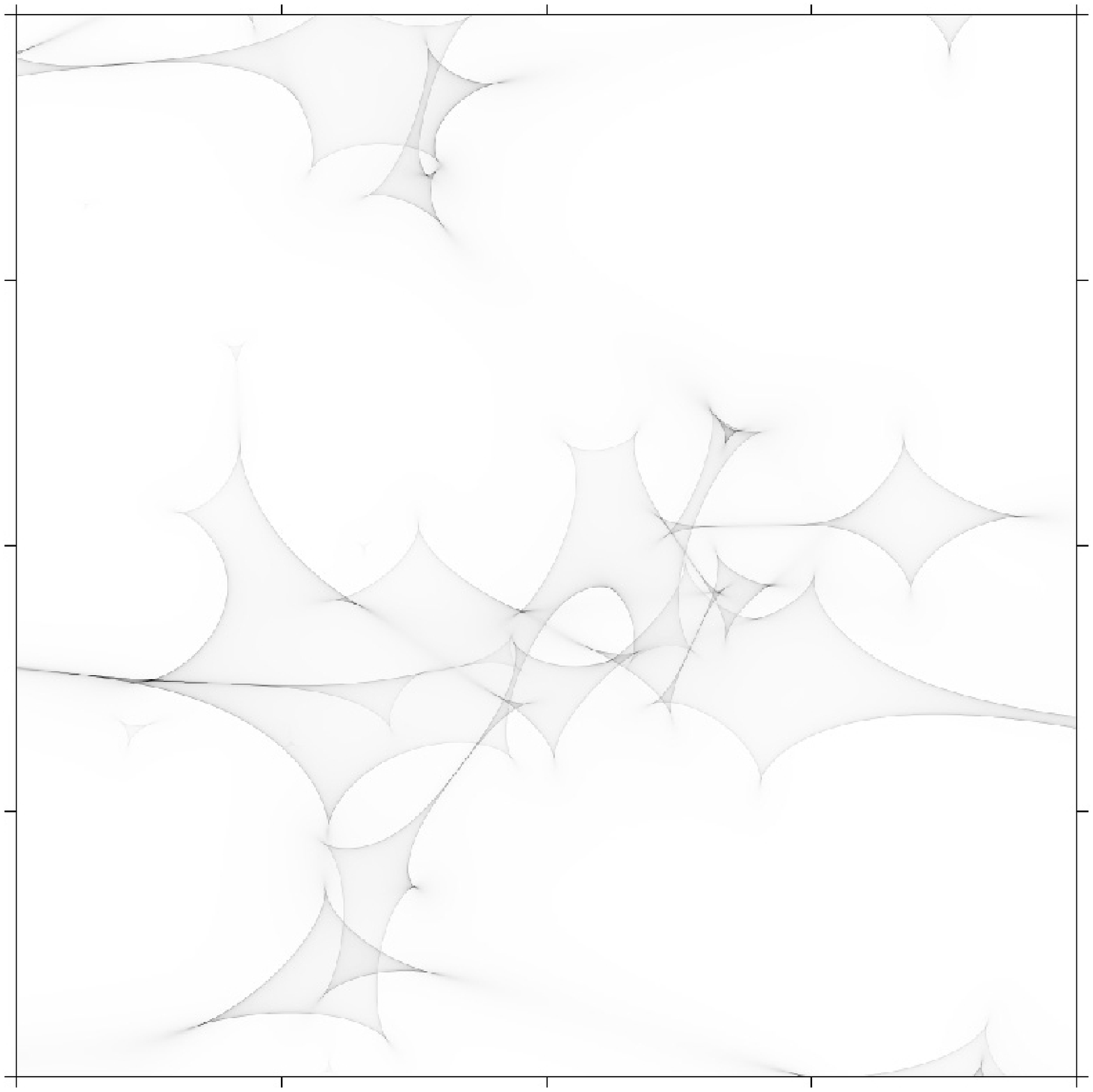}
\includegraphics[angle=0,width=\textwidth]{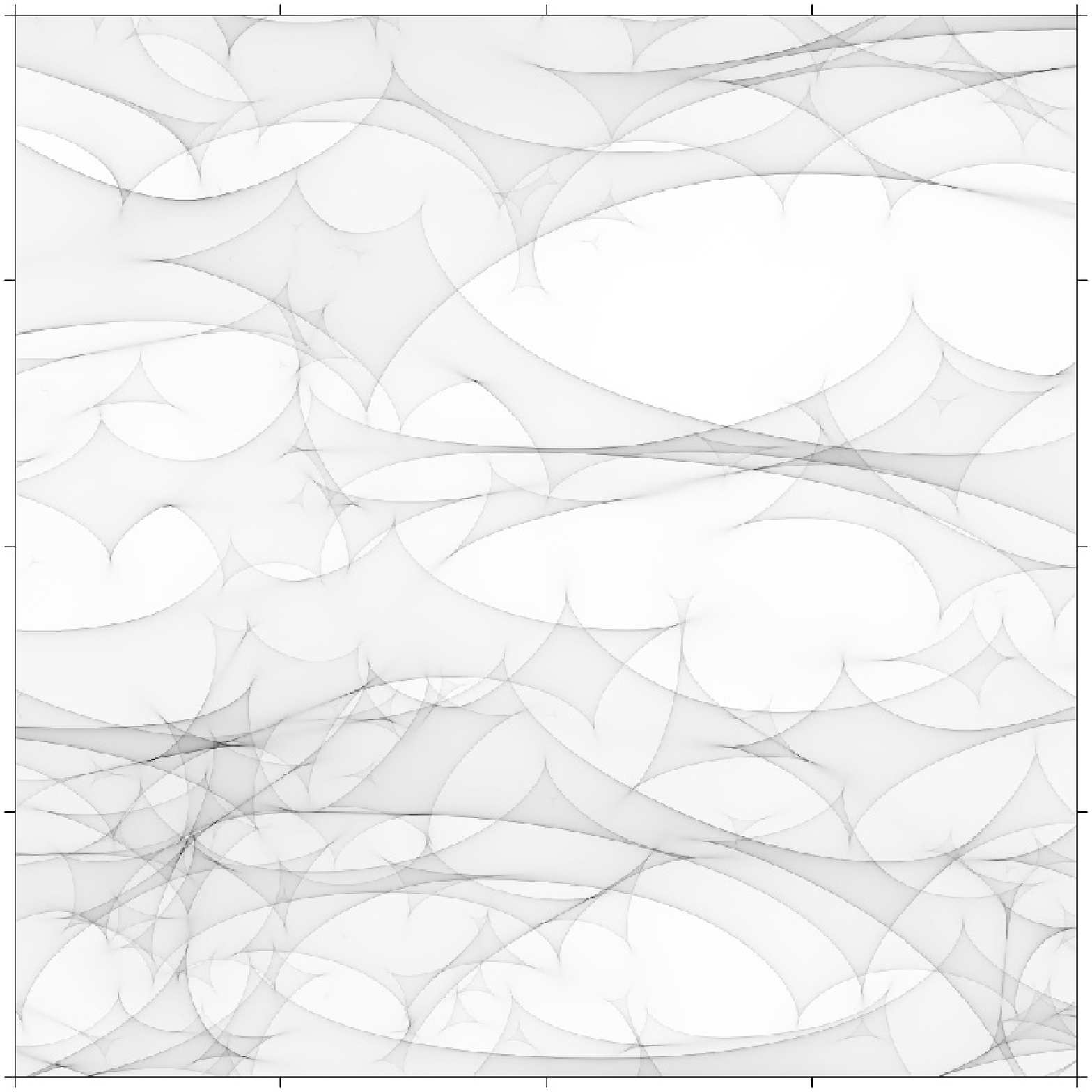}
\end{minipage}
\begin{minipage}[c]{0.25\textwidth}
\includegraphics[angle=0,width=\textwidth]{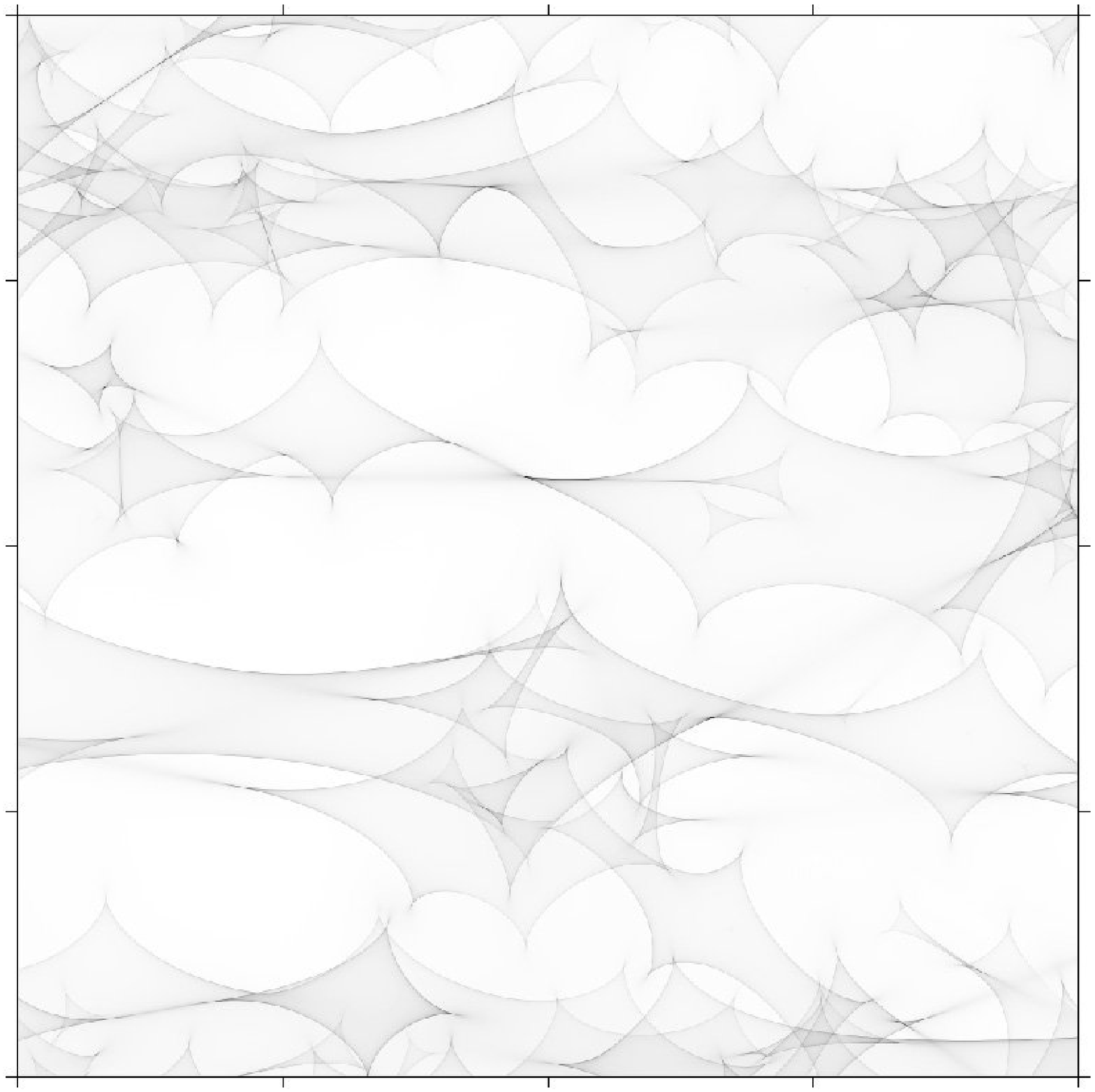}
\includegraphics[angle=0,width=\textwidth]{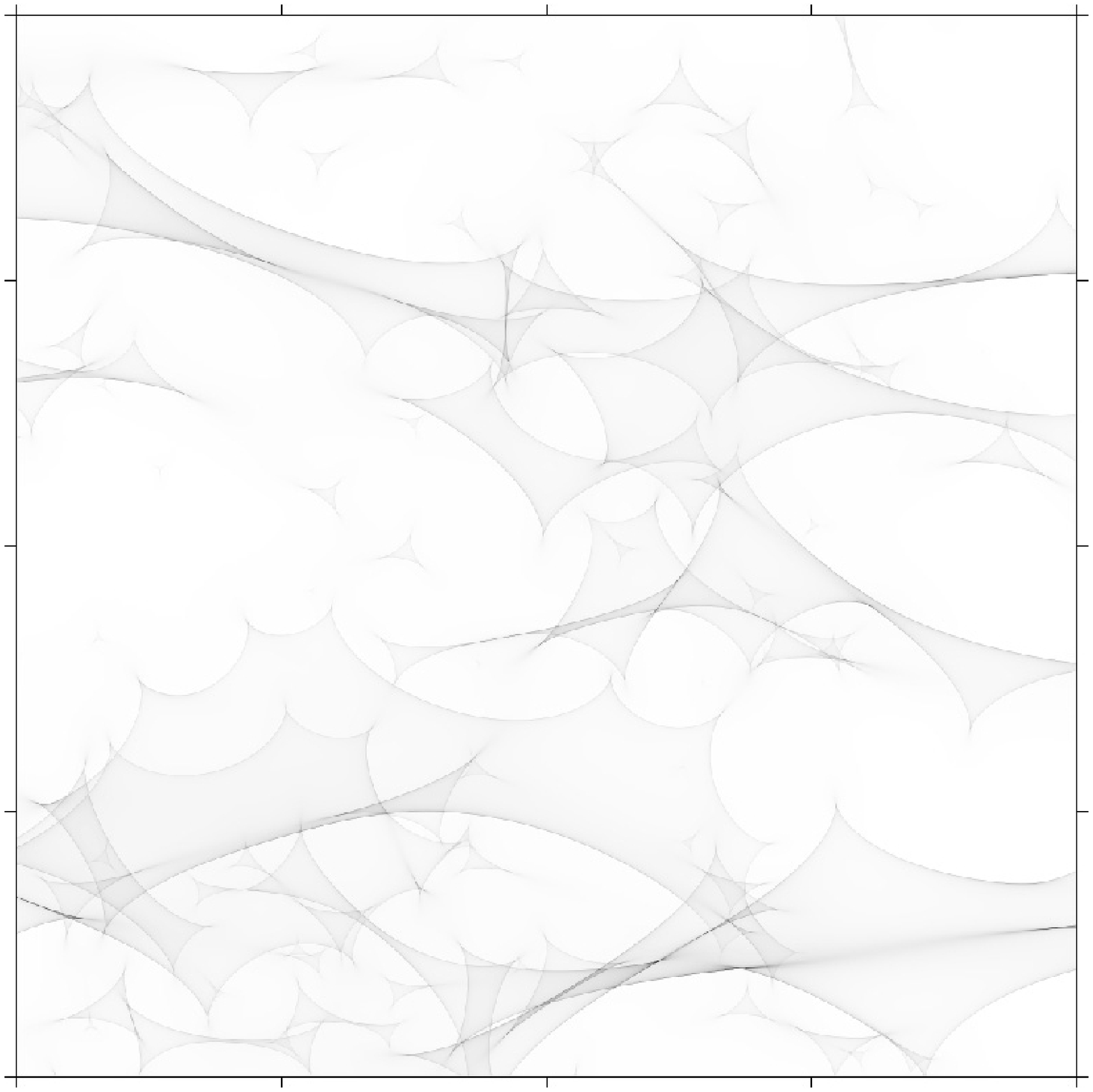}
\end{minipage}
\caption{Example magnification  maps for $\sigma=\gamma=0.25$  (top left), 0.4
  (top right), 0.6  (bottom left), and 0.75  (bottom right). Darker regions  indicate higher
  magnification, with the sharp boundaries denoting caustics of the map. 
  Each have dimension  10x10 ER.}\label{fig5}
\end{center}
\end{figure*}

There are  two main  parameters required to  model the microlensing  which are
dependent upon the mass distribution in the lensing galaxy - the dimensionless
surface mass density $\sigma$ (or optical depth) and the shear $\gamma$ due to
the large  scale matter  distribution.  As these  parameters are  not strongly
constrained  for  the Cloverleaf,  this  study  has  investigated 4  different
models: $\sigma=\gamma=0.25$,  0.4, 0.6 and  0.75.  These were  chosen because
they  represent a  singular  isothermal sphere,  reasonably approximating  the
range  of  possible mass  distributions  for  a  lensing galaxy  \citep{schn}.
Figure \ref{fig5}  shows example magnification maps for  regions derived using
the ray tracing method, with side lengths of 10 ER for the 4 cases of $\sigma$
and $\gamma$ above.

These  maps can  be  scaled to  physical  distance using  the Einstein  radius
(Equation \ref{eq2}).  Assuming a  standard cosmological model  with $h=0.73$,
$\Lambda=0.76$  and  $\Omega=0.24$  \citep{spe},   and  the  lens  and  source
redshifts given in Section~\ref{h1413}, ER$\approx$2.73x$10^{16}$ cm.

The  magnifications due to  microlensing  fluctuate  about a  mean theoretical
value given by
\begin{equation}
\mu_{th}=\left[ \left( 1-\sigma^{2} \right) - \gamma^{2} \right]^{-1}\label{eq3}
\end{equation}
which is the magnification an image would suffer if the macrolens was composed 
of solely smooth matter. Hence, generally, the mean magnification over a large 
enough area should tend to this theoretically expected value. However, due to
statistical variance, the mean value within an individual magnification map 
which is relatively small can deviate from this expected value (i.e. the smaller
the region chosen, the larger the possible deviation from the mean theoretical
magnification).

\subsection{Defining the Source} 
\begin{figure*}
\begin{center}
\includegraphics[scale=0.6, angle=0]{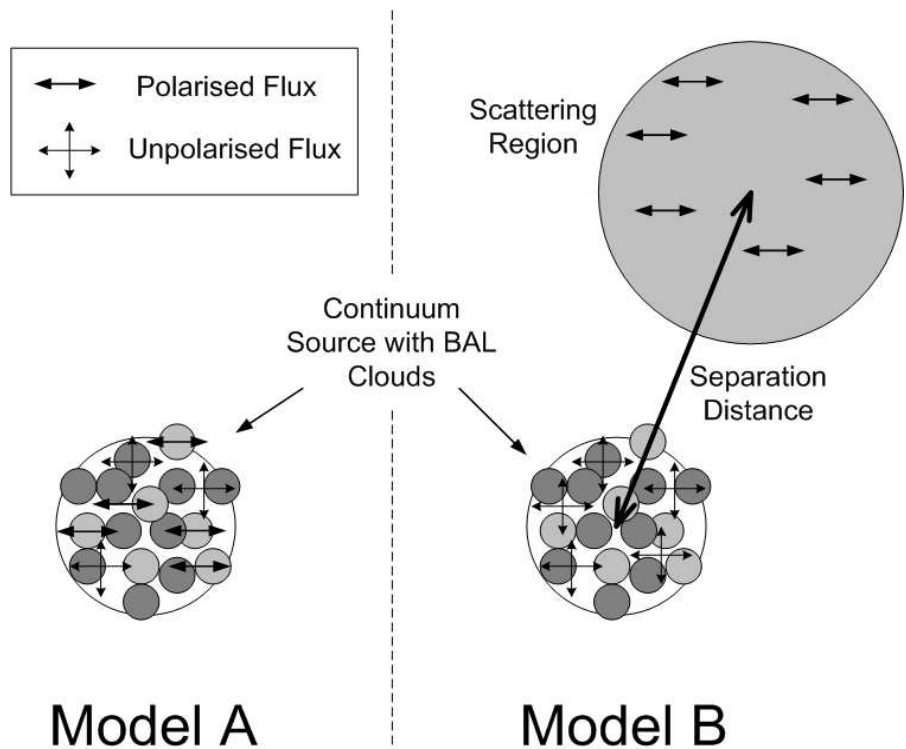}
\caption{Schematic view of the BAL quasar continuum and scattering region as seen by the
  microlensing magnification for both scattering  scenarios. This is based on Figure
  \ref{fig2}   and   is  discussed   further   in   Sections  \ref{modA}   and
  \ref{modB}.}\label{fig6}
\end{center}
\end{figure*}

In order  to determine the effect  of microlensing on each  model from Section
\ref{h1413}, it was necessary to construct the image that the magnification map
would `see'. This is shown  schematically  in Figure  \ref{fig6} and  discussed
in detail below.

\subsubsection{Absorption and Polarisation}
In order to examine  a BAL trough, Equation \ref{eq4} was used to simulate the
blue wing of C$_{\textrm{IV}}$ $\lambda$1549, where $\lambda_{cc}=1505$ \AA\ is
the wavelength at which 50\% of  the light from the continuum core is absorbed
(corresponding to  a bulk outflow at  50\% absorption of  $\sim$8500 \kms) and
2$\Delta\lambda=8.75$ \AA\ is the  `width' of  the absorption  feature between
27-73\% absorption \citep{tur:2,wey}.
\begin{equation}
A\left(\lambda\right)=\frac{1}{1 + exp\left[\left(\lambda-\lambda_{cc}\right)/\Delta\lambda\right]}\label{eq4}
\end{equation}
With  regard to  polarisation, as  discussed  in the  following sections,  the
polarisation  at any  pixel was  directly related  to the  absorption  at that
pixel. This could take on any value in the range from  0-20\%. This would tend
to generate  conservative results for this  study (note that,  as mentioned in
Section \ref{h1413}, the  polarisation of the entire composite  image could be
as  high as 20\%);  however, any  general trends  established here would still
apply if a larger range for polarisation was used.

\subsubsection{Model A: Scattering Within the BAL Region}\label{modA}
For this study  the quasar continuum source was  modelled as a two-dimensional
Gaussian surface with a brightness radius of $10^{15}$ cm \citep{ree,bla} on a
64x64 pixel grid. The extent of this grid was defined so that the value of the
Gaussian at the edges would be 10\% or less than the value at the peak, giving
a side length of 3.65x$10^{15}$ cm (this  kept  the  size  of  the  grid  down
while  allowing for  good  resolution). Hence,  this  Gaussian represents  the
unabsorbed continuum flux and in  order to represent absorption, an absorption
matrix (discussed shortly) would then be (dot) multiplied with the source grid
(i.e.  this absorption matrix is the same size as the source matrix).

In  order to  represent the  scale length  of the  inhomogeneities in  the BAL
region  (this will be  referred to  as cloud  size), cloud  sizes were  set as
3.65x$10^{15}$x$4^{-n}$  cm, n=1,2,3. In  this way,  the smallest  clouds were
5.70x$10^{13}$ cm in extent (n=3) while the largest clouds were 9.11x$10^{14}$
cm (n=1).   This was implemented by  limiting the initial matrix  size for the
absorption matrix generator  and then drawing this matrix out  to fill a 64x64
grid. Note, this method was similar to that employed by \citet{lew:1}.

\begin{figure*}
\begin{center}
\begin{minipage}[c]{0.31\textwidth}
\includegraphics[angle=0,width=\textwidth]{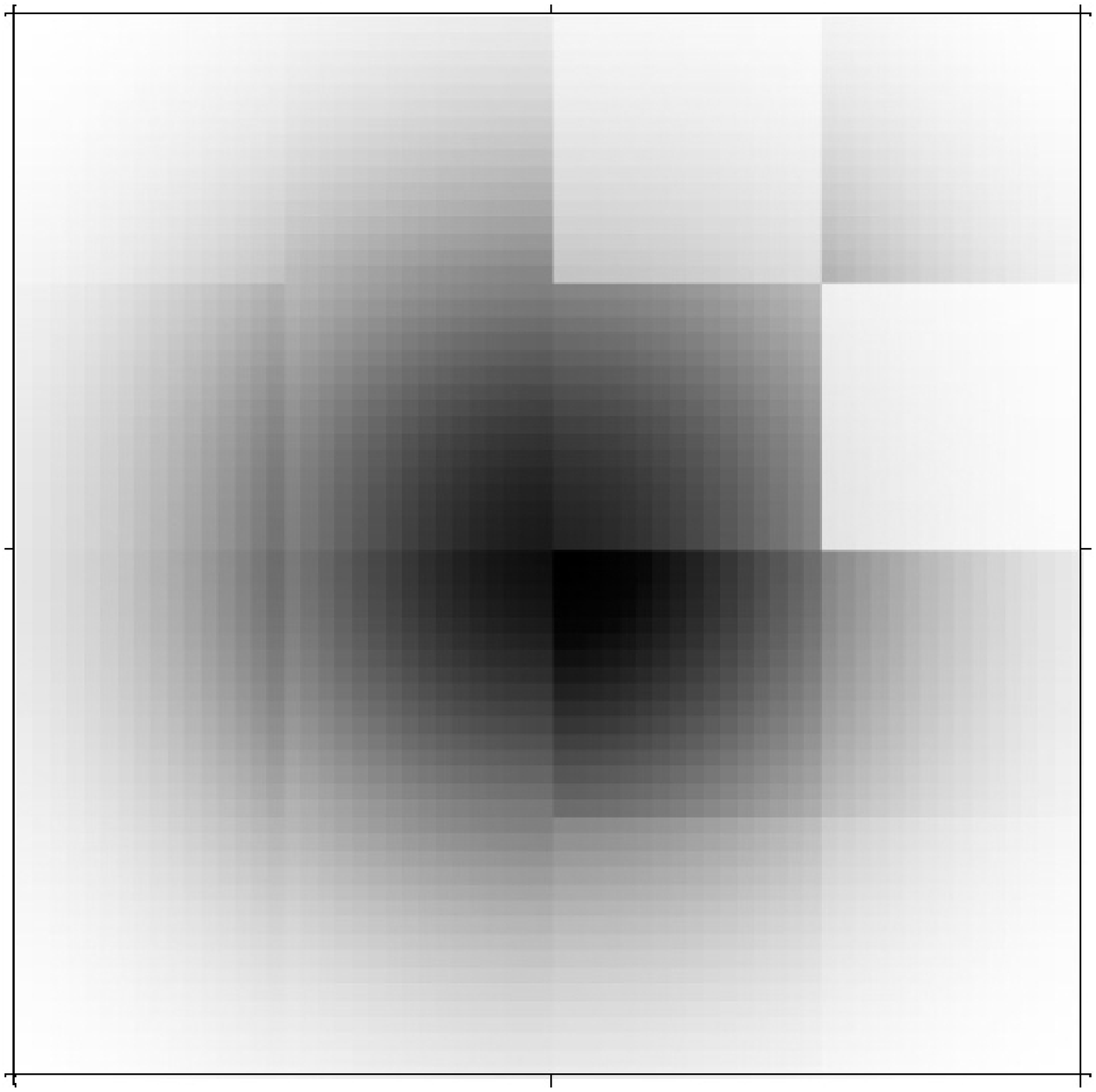}
\end{minipage}
\begin{minipage}[c]{0.31\textwidth}
\includegraphics[angle=0,width=\textwidth]{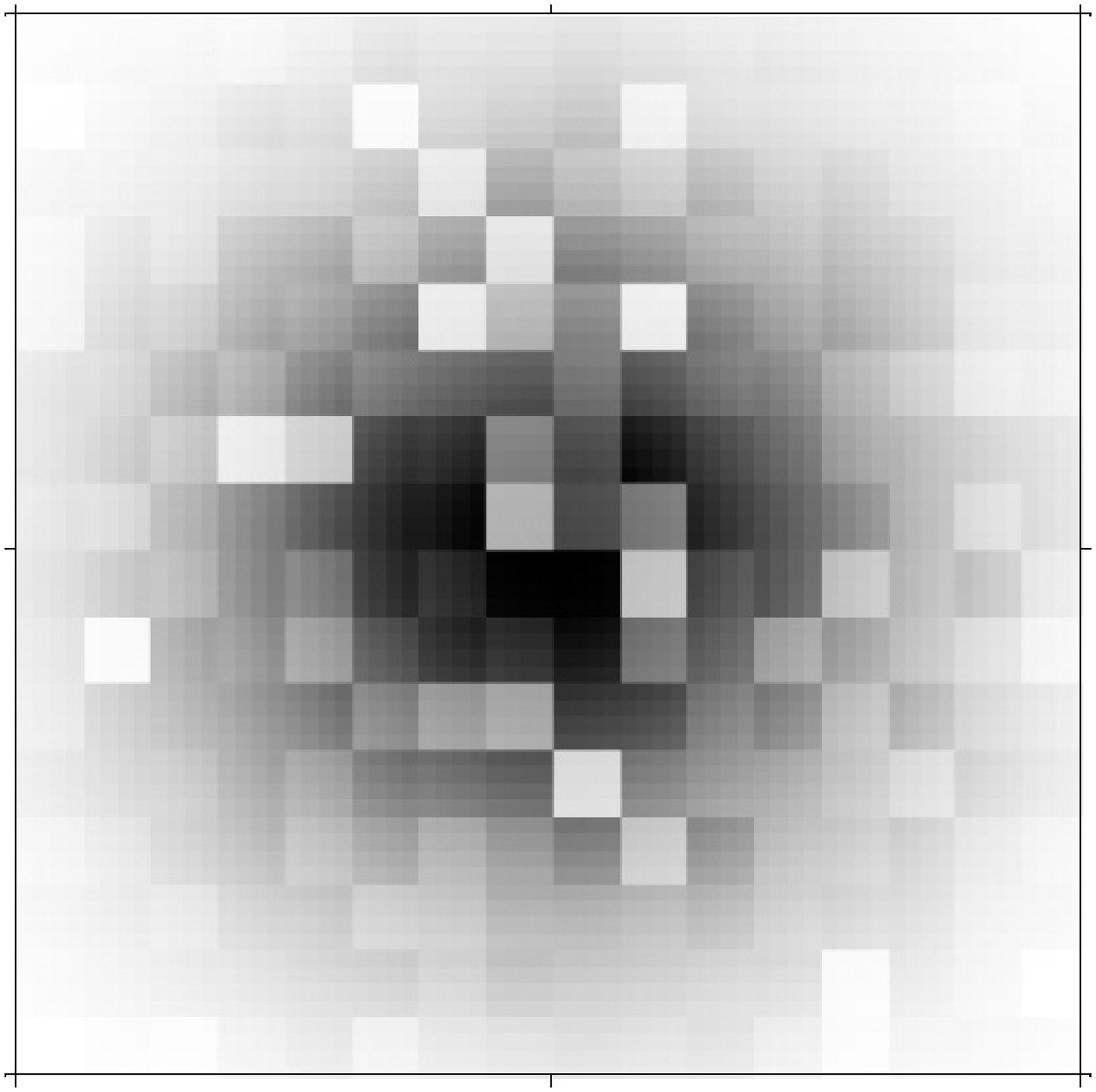}
\end{minipage}
\begin{minipage}[c]{0.31\textwidth}
\includegraphics[angle=0,width=\textwidth]{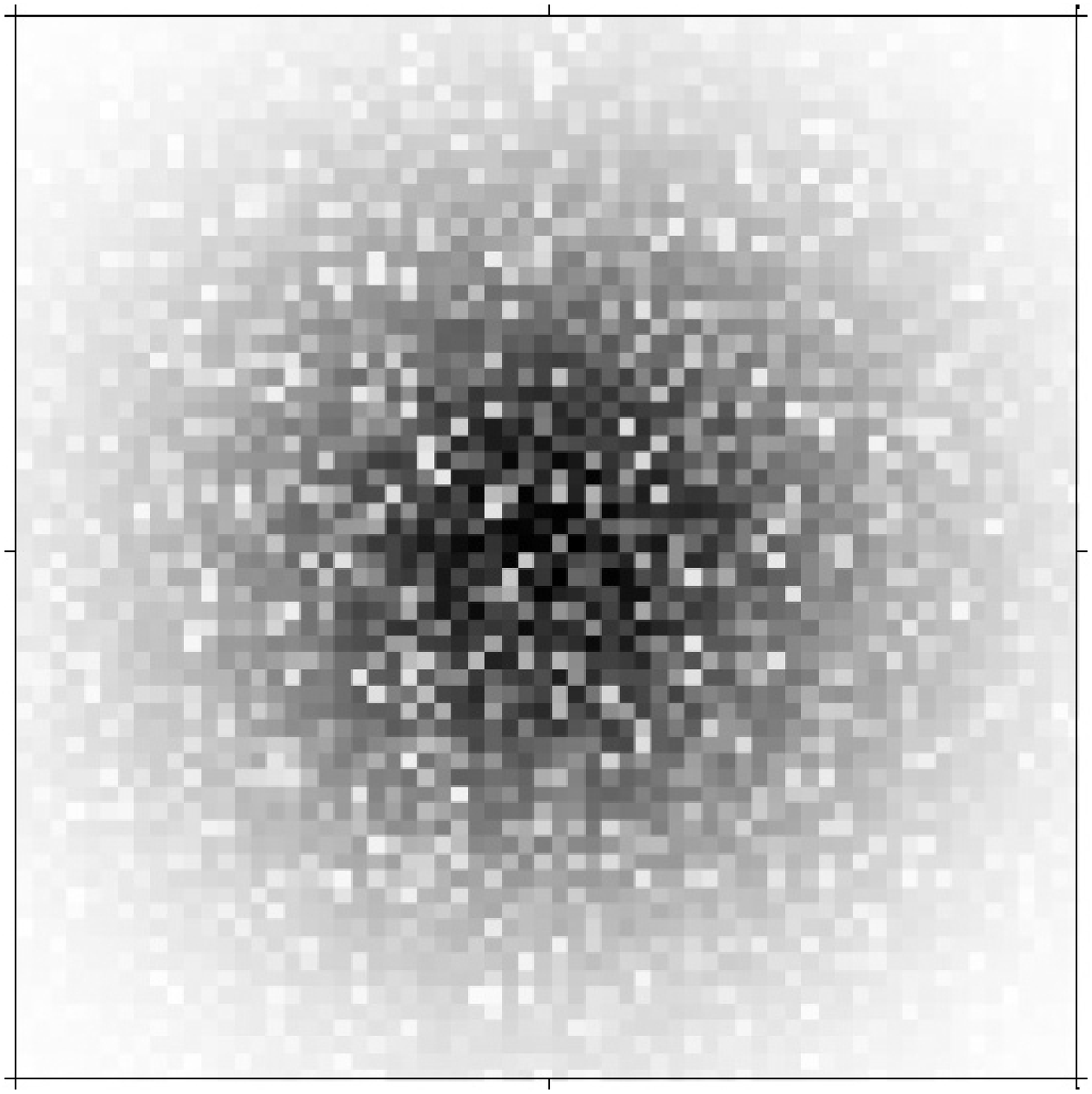}
\end{minipage}
\caption{Model  A.  Left:  Source   with  cloudsize  9.11x$10^{14}$  cm  (n=1),
  absorption at 25\% and $\omega$  at 50\%.  Middle and Right: Same parameters
  but with cloudsize  2.28x$10^{14}$ cm (n=2) and 5.70x$10^{13}$  cm (n=3). Each
  have side length 3.65x$10^{15}$ cm.}\label{fig8}
\end{center}
\end{figure*}

The absorption matrix represents the degree of variation of absorption
about a particular  wavelength (i.e.  due to inhomogeneities  in the density  of the
absorption region). The  distribution of values were represented  by a Gaussian
of specified width  $\omega$ centred at  $A_{c}(\lambda)$.
A modification  algorithm was developed to truncate this distribution
outside 0-100\%, and to reassign new values for the truncated elements
such that the overall distribution of matrix values had mean
$\left<A\left(\lambda\right)\right>$ and width $\omega$. The mean of the
absorption distribution $\left<A\left(\lambda\right)\right>$ was chosen
to be a function of wavelength as given by Equation \ref{eq4}.
The algorithm    was   also   designed    so   that    if,   for    a   particular
$\left<A\left(\lambda\right)\right>$   and   $\omega$,  $A_{c}(\lambda)$   was
located outside the 0-100\% boundary,  then $A_{c}(\lambda)$ would be fixed at
the    boundary    and   $\omega$    varied    so    as    to   ensure    that
$\left<A\left(\lambda\right)\right>$  was the value specified. This allows, for example,
for a distribution with (mean) absorption of 10\% and width of 50\% to be generated so that
all matrix values lie within 0-100\%. For  this
paper, widths of  5\% and 50\% were investigated for  absorption at 25\%, 50\%
and 75\%.  Figure \ref{fig8} shows how different cloud sizes were modelled
with a particular set of absorption matrix parameters.

In  order  to implement  polarisation,  this image  was  then  split into  two
separate images  according to  Equation \ref{eq5}.  In  this way,  if Equation
\ref{eq6}  was  calculated  at  each  pixel then  the  polarisation  would  be
dependent  on the  particular absorption  matrix  value at  that pixel.  Here,
$Pix_{n}$  are  the  values  at   corresponding  pixels  in  each  new  image,
$Pol_{unab}=0.02$, $Pol_{lim}=0.18$, $Abs$  is the absorption matrix value
at  that pixel,  and  $Pol$ is  the  polarisation difference  between the  two
images.  In this  way, if  $Abs$=0\%  then $Pol$=2\%,  if $Abs$=50\%  then
$Pol$=11\%, and if $Abs$=100\% then $Pol$=20\%.

\begin{eqnarray}
Pix_{1}=0.5 P_{tot} \left( 1 + Pol_{unab} + Pol_{lim} \times Abs\right) \nonumber \\
Pix_{2}=0.5 P_{tot} \left( 1 - Pol_{unab} - Pol_{lim} \times Abs\right) \nonumber \\
P_{tot}=Pix_{1} + Pix_{2} \nonumber \\
P_{dif}=Pix_{1} - Pix_{2} \label{eq5}
\end{eqnarray}

\begin{equation}
Pol=\frac{P_{dif}}{P_{tot}}\label{eq6}
\end{equation}

\subsubsection{Model B: External Scattering Region}\label{modB}
This  model was  somewhat constrained  by the  maximum magnification  map that
could be handled  in terms of computer  memory, so by setting this  map with a
side  length  of   100  ER  (1024x1024  pixels)  the   pixel  scale  size  was
automatically set  for any  source models created.   Fortunately this  did not
seriously affect the range of scattering  region scale sizes that needed to be
investigated.  Unlike Model A, an absorption matrix was not required here (the
continuum  region is  on the  order of  a few  pixels), but  rather absorption
simply involved reducing  the flux of the continuum while  keeping the flux of
the scattering region constant.

\begin{figure*}
\begin{center}
\begin{minipage}[c]{0.31\textwidth}
\includegraphics[angle=0,width=\textwidth]{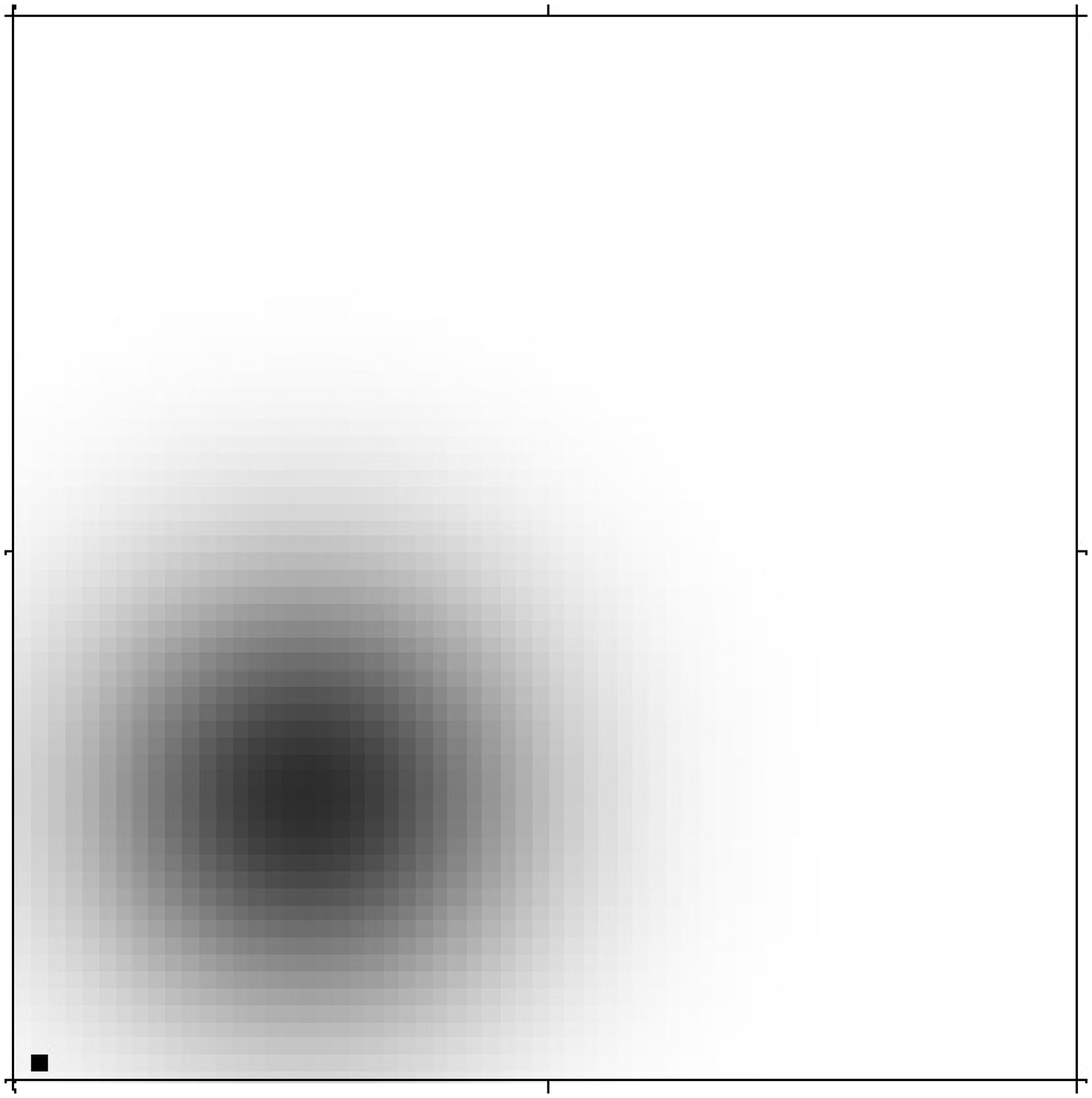}
\end{minipage}
\begin{minipage}[c]{0.31\textwidth}
\includegraphics[angle=0,width=\textwidth]{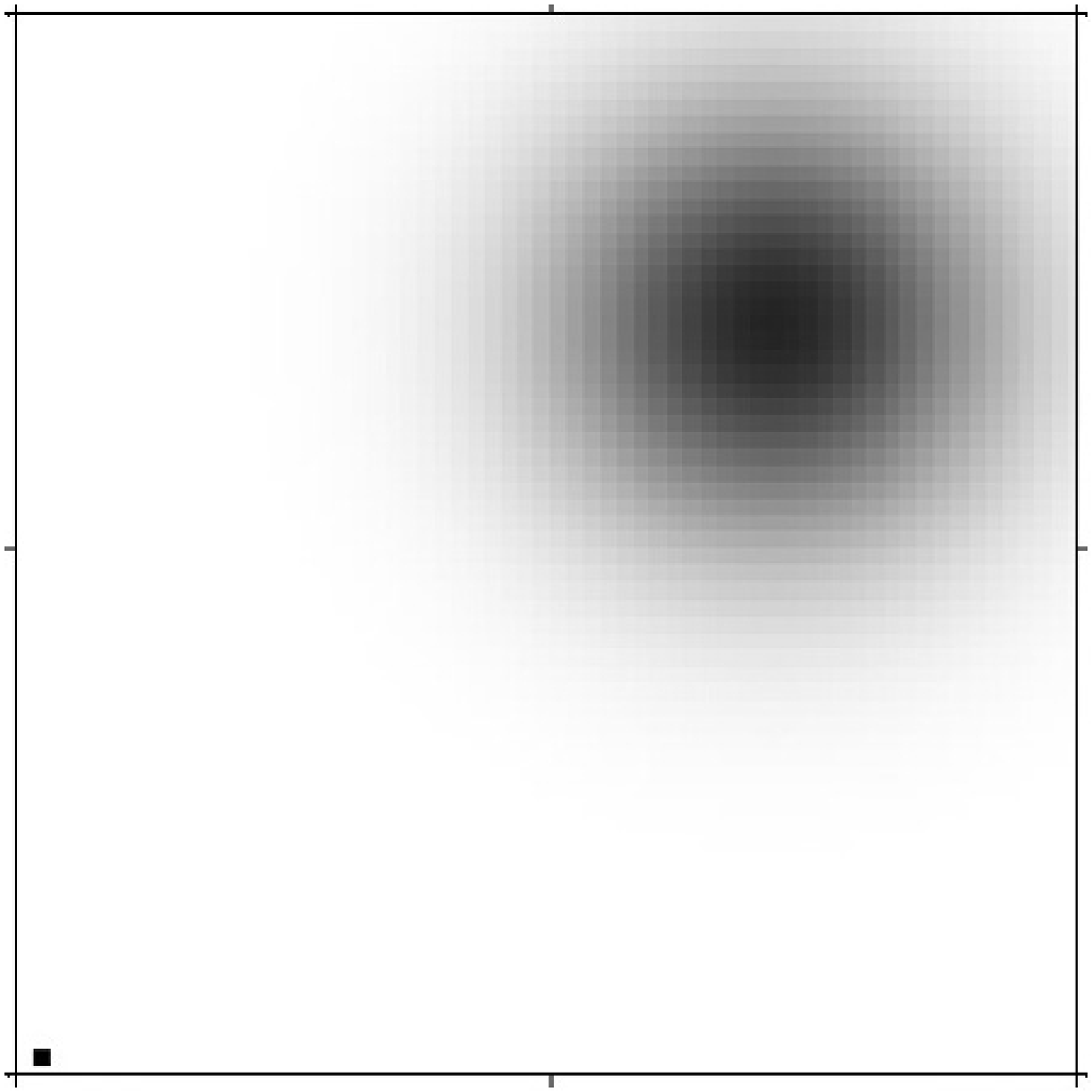}
\end{minipage} \\
\begin{minipage}[c]{0.31\textwidth}
\includegraphics[angle=0,width=\textwidth]{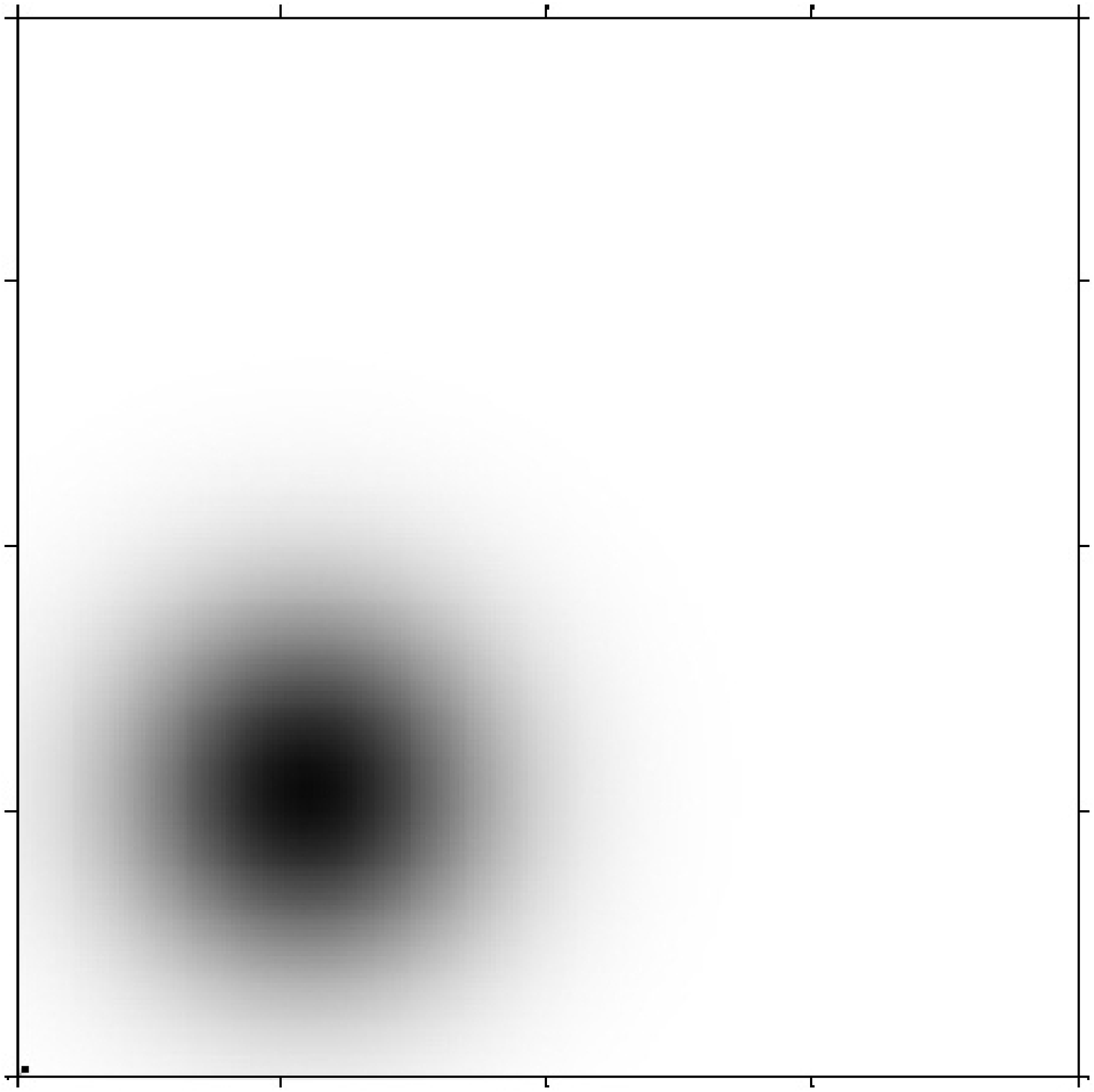}
\end{minipage}
\begin{minipage}[c]{0.31\textwidth}
\includegraphics[angle=0,width=\textwidth]{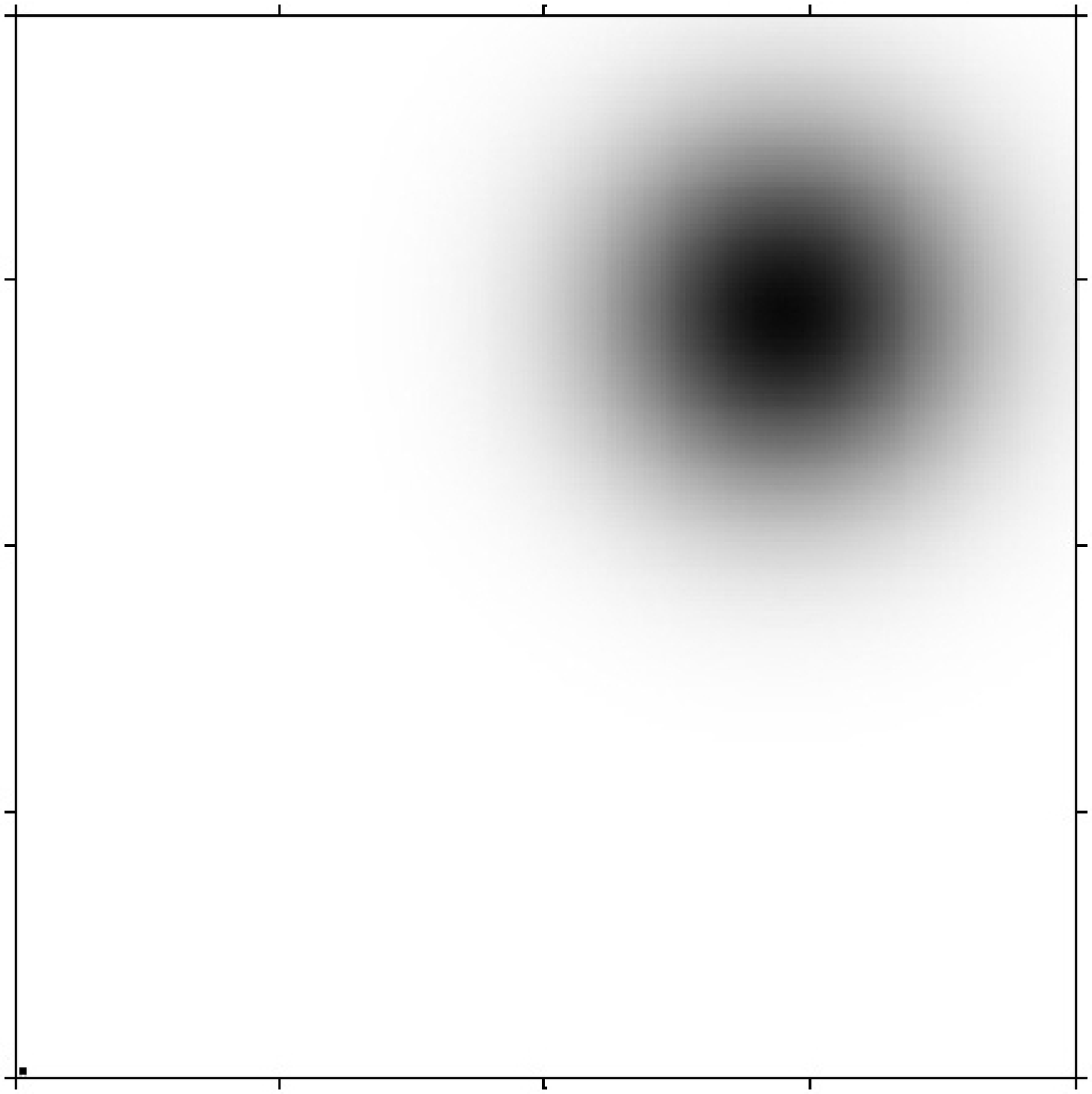}
\end{minipage}
\begin{minipage}[c]{0.31\textwidth}
\includegraphics[angle=0,width=\textwidth]{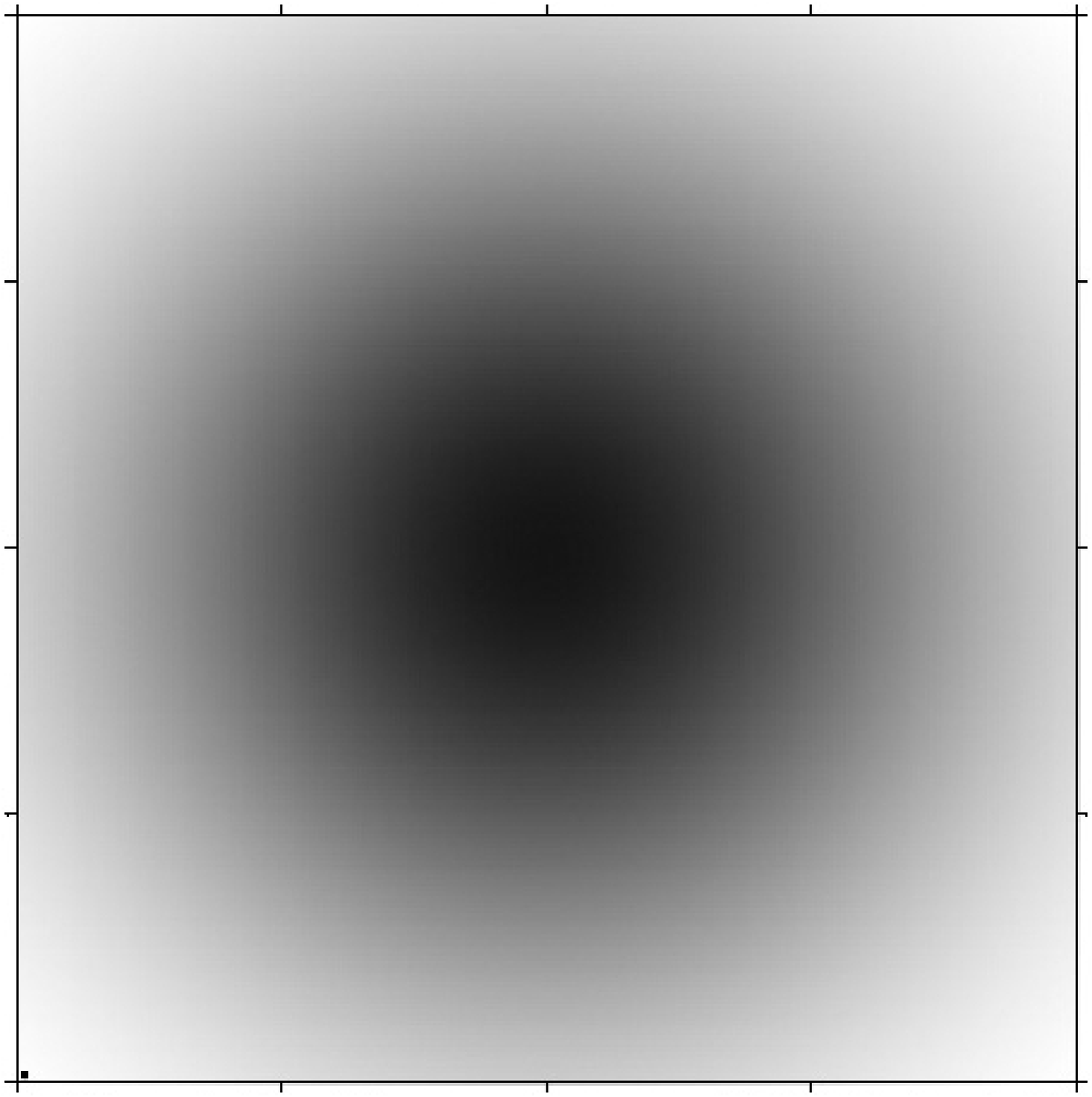}
\end{minipage}
\caption{This shows the  5 different sources considered as part  of Model B at
  high absorption  percentages (so  that the continuum  doesn't drown  out the
  scattering region in the images). Scattering  region  radii are 3.00x$10^{16}$
  cm with  separation distances of  6.03x$10^{16}$ (top left) and 1.66x$10^{17}$
  cm (top right),  7.70x$10^{16}$  cm  with  separation  distances  of
  2.00x$10^{17}$ (bottom left)  and  5.39x$10^{17}$ cm  (bottom middle),  and
  2.00x$10^{17}$  cm with  a  separation distance  of  3.70x$10^{17}$ cm  (bottom
  right). Images in the top row have side length 1.71x$10^{17}$ cm whereas images
  in the bottom row have side length 5.34x$10^{17}$ cm. Note  the continuum in
  the  bottom left corners.}\label{fig9}
\end{center}
\end{figure*}

This left two main variables  to manipulate besides absorption: the scale size
of the scattering  region and the separation distance of  this region from the
continuum. Note that  the latter is in fact intrinsically  linked to the scale
size of  the scattering region  in that at  scale sizes approaching  the upper
limit, the separation  distance must be of the order  of the scattering region
radius (geometries where  the separation distance is less  than the scattering
region radius  would give similar  correlation results). The scattering region
radii and separation distances investigated for this model are shown in Figure
\ref{fig9}. This involved setting  the source as a 64x64 grid for the smallest
scattering region and 200x200  for the larger  two  regions (resulting in much
higher  computation  times). Also  note  that, like  the  continuum  region in
Model A, both the continuum and  scattering  regions in  Model B were modelled
using two-dimensional Gaussian surfaces.

The method for achieving polarisation in this model is quite different to that
of Model  A. Here,  polarisation is due  to the  reduced flux coming  from the
continuum so  that, unlike Model A,  a linear increase in  absorption will not
create a linear  increase in polarisation. For this  reason, absorption values
of  77\%,  91\%  and 97\%  were  chosen  so  as  to  represent the  same  mean
polarisation values  used in Model A.  To generate this  source, the continuum
region flux  is diminished  in accordance with  the absorption value  and then
split between two images. The scattering region is then added to each image so
that they satisfy Equation \ref{eq6}  for a constant polarisation of 20\%.  In
addition,  the relative  flux  of  the two  sources  can then  be  set by  the
requirement  that  the  polarisation  at  0\% absorption  is  2\%  (this  then
increases non-linearly to 20\% polarisation at 100\% absorption).

\subsection{Convolution}\label{conv}
In order  to determine the  magnification distributions of the  various source
images,  these images need  to be  convolved with  the magnification  maps. In
doing this it was important to ensure that the physical scale corresponding to
the pixel length of both source and magnification map was identical. For Model
A this was  determined by the design of the  source, requiring a magnification
map of 2.136x2.136 ER (1024x1024 pixels). As this region is  reasonably  small
it was important to recognise that a number of magnification maps would need to be
used to obtain data (see Section \ref{raytrac}).  Neglecting this could reveal
incorrect trends  in any correlation data  (in particular it  would reduce the
range of  magnifications significantly). For  this study, two maps  of varying
mean magnification were used to ensure  that possible trends at both large and
small magnification would not be overlooked (e.g. a trend at low magnification
might be  overlooked if dealing  with a magnification map  excessively covered
with  intense  caustic  structures);  this  is discussed  further  in  Section
\ref{resA}. Note that this problem does  not exist for Model B because the map
is large enough that,  on the whole, it is reasonably homogeneous  and so has a
mean  that  approaches  the   theoretical  expectation  value.   As  mentioned
previously, the process for Model B was the reverse of that for Model A, whereby
the largest magnification map of 100x100 ER was created in order for the source
to then integrate with it.

In order  to determine the  polarisation following convolution (using  the two
polarised images formed  for Model A or B) Equation  \ref{eq6} was used. These
values could then be compared  directly with the magnification values obtained
from convolving the magnification  map with an unabsorbed, unpolarised version
of  the  corresponding  source  (these  values were  divided  by the mean flux
of the original  source in order to obtain  the correct magnification values).
Correlation maps could then be obtained. Finally, note that for this paper all
convolutions  were  computed without  any  zero-padded  edges, so  that  areas
contaminated by the overlaps were neglected.

\section{Results}
\begin{figure*}
\begin{center}
\begin{minipage}[c]{0.2\textwidth}
\includegraphics[angle=0,width=\textwidth]{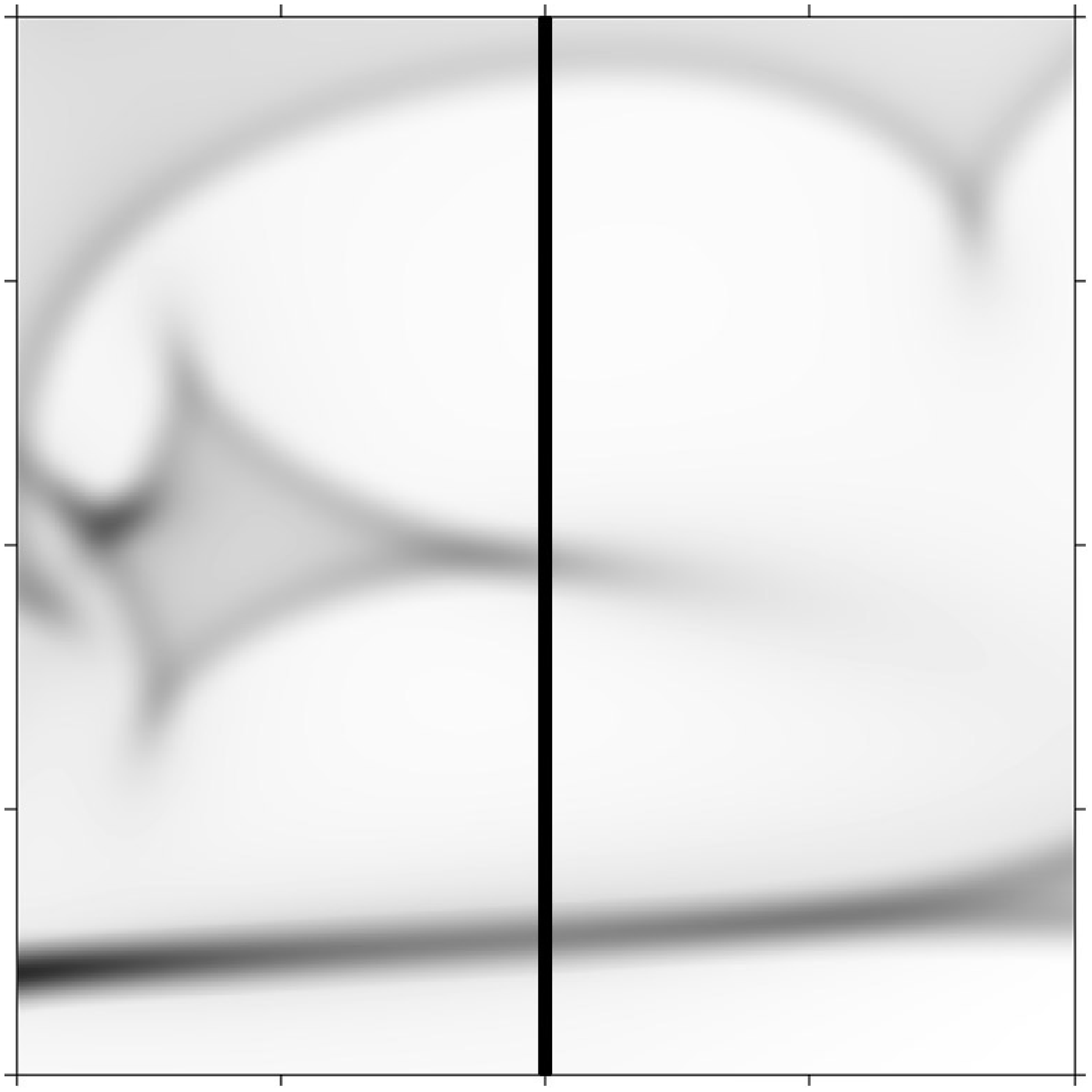}
\end{minipage}
\begin{minipage}[c]{0.2\textwidth}
\includegraphics[angle=0,width=\textwidth]{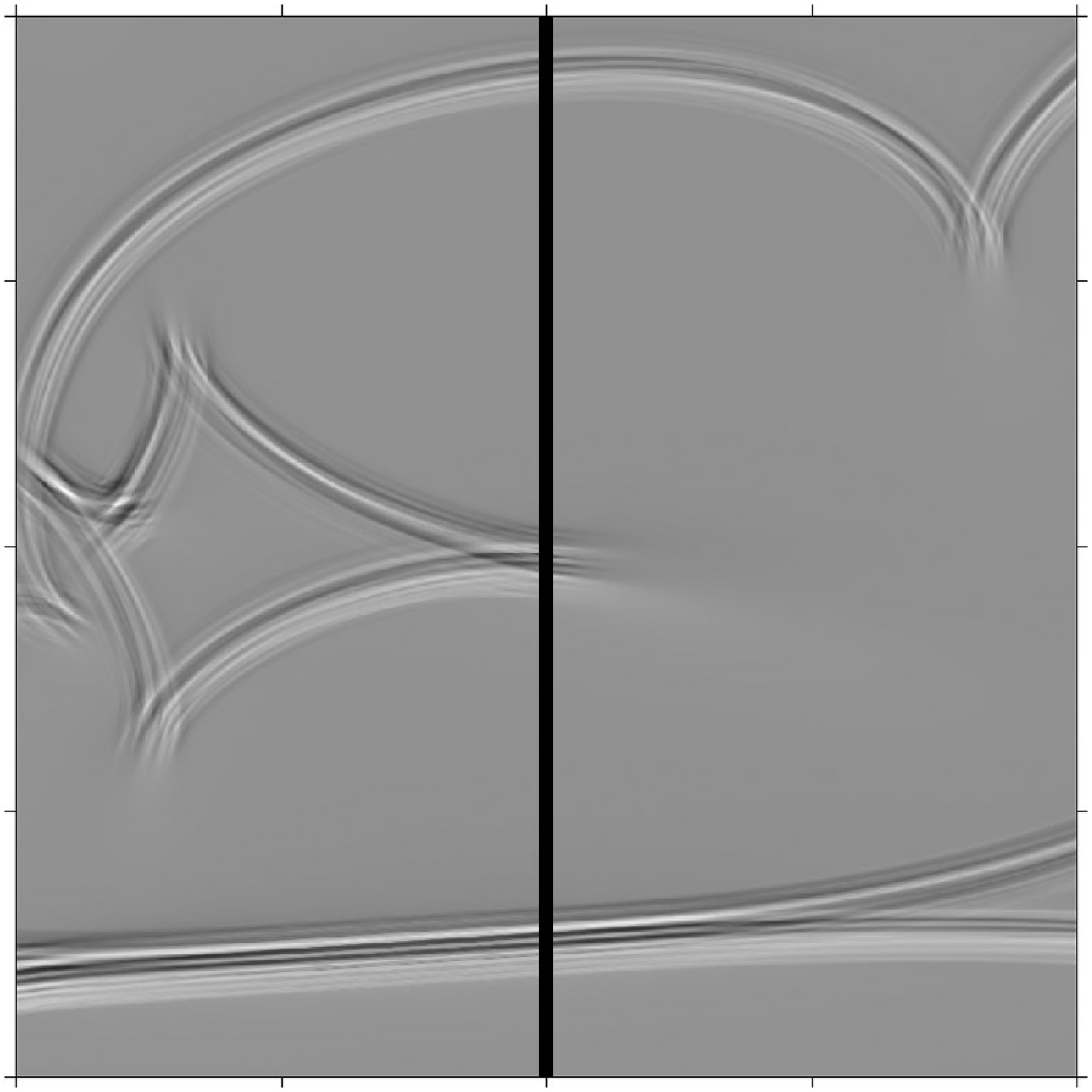}
\end{minipage}
\begin{minipage}[c]{0.5\textwidth}
\includegraphics[angle=0,width=\textwidth]{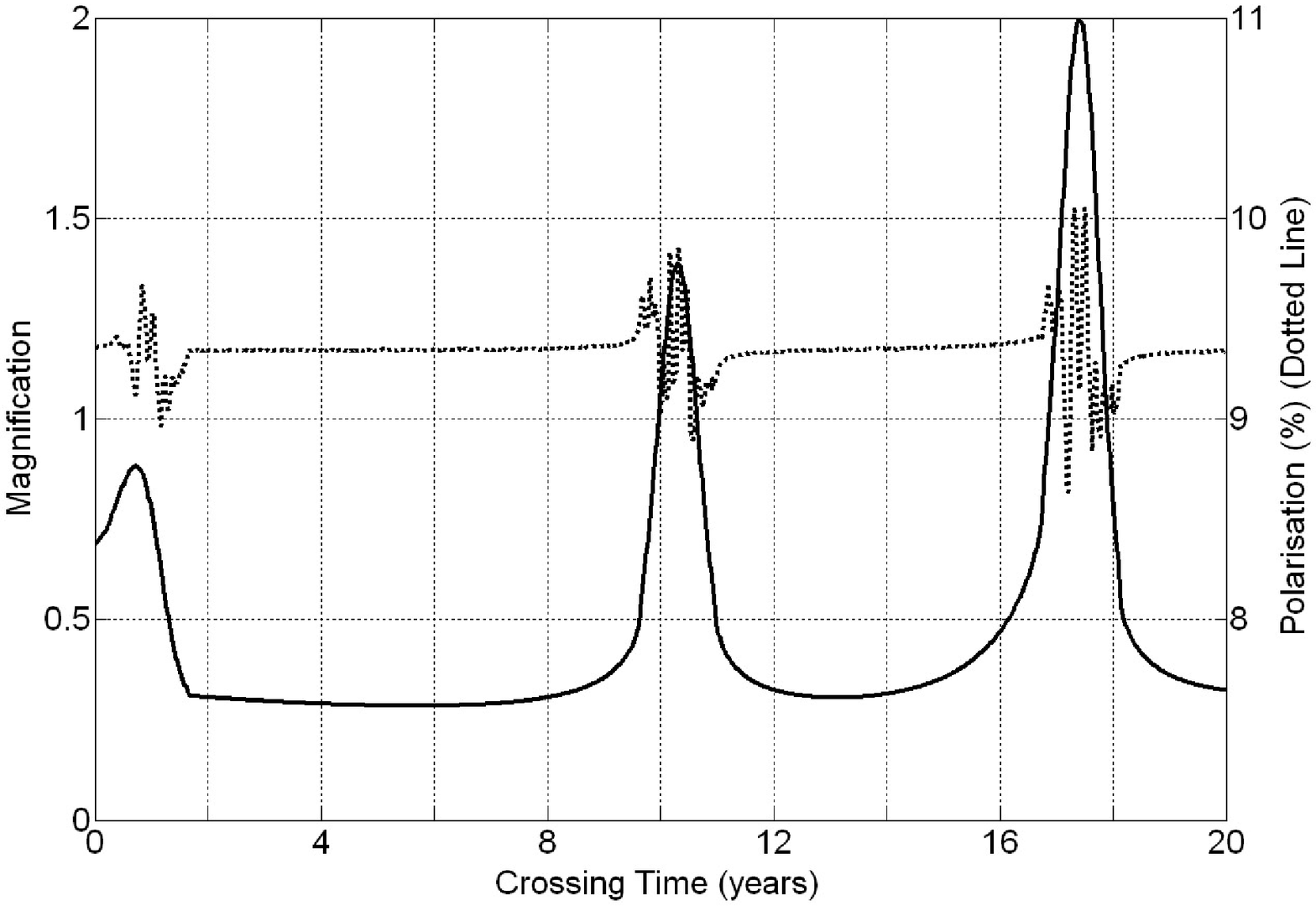}
\end{minipage}
\caption{Model  A.  Left:  Convolution  of  unabsorbed source  with  2.136x2.136  ER
  magnification  map  ($\sigma=\gamma=0.4$).   Middle: Resultant  polarisation
  after application of Equation \ref{eq6} to two convolutions of the polarised
  source (absorption=50\%, width=50\%).  Both have side length 5.47x$10^{16}$ cm. Right: Light curves for magnification
  and  polarisation, as  indicated  by line  from  top to  bottom in  previous
  images.}\label{fig10}
\end{center}
\end{figure*}

\begin{figure*}
\begin{center}
\begin{minipage}[c]{0.2\textwidth}
\includegraphics[angle=0,width=\textwidth]{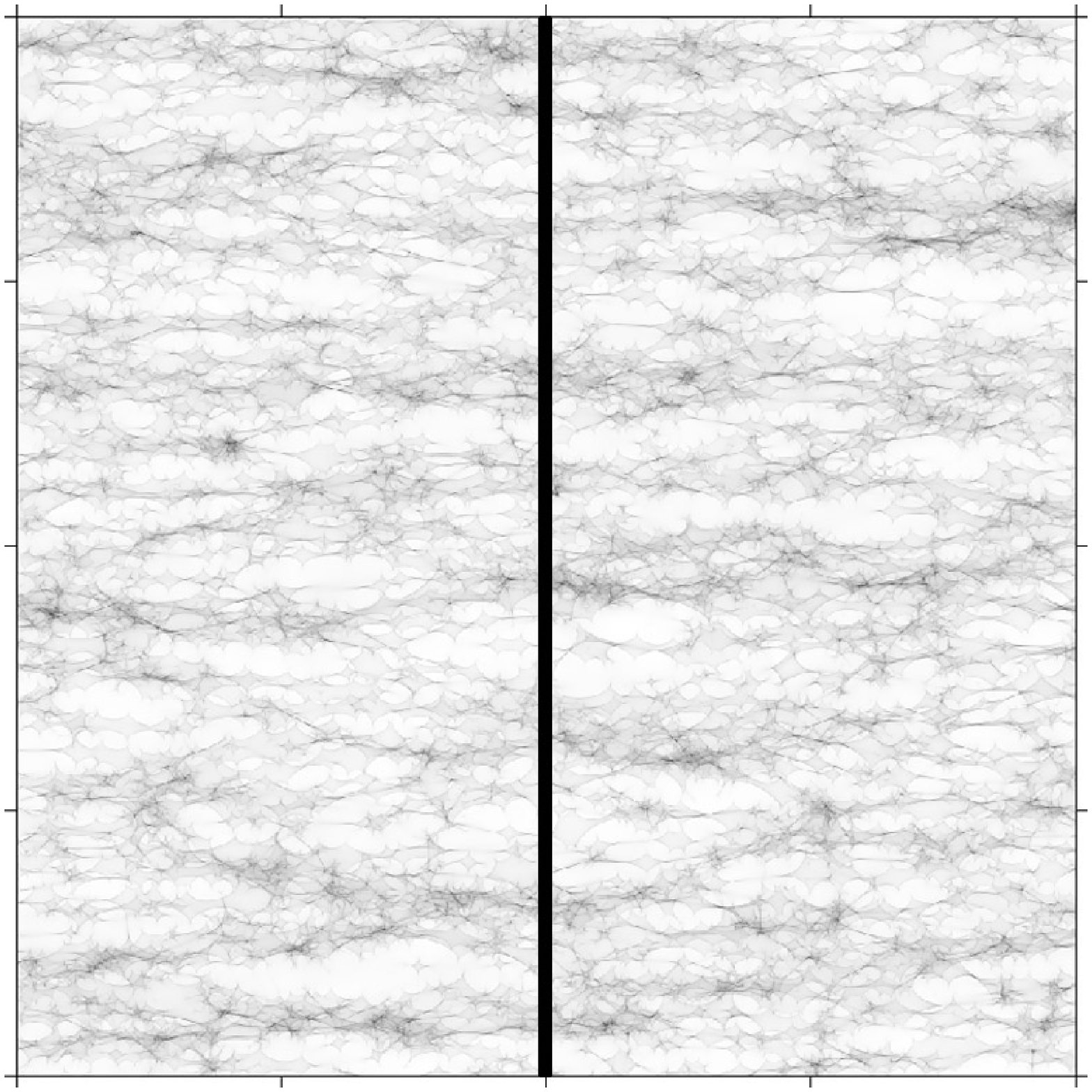}
\end{minipage}
\begin{minipage}[c]{0.2\textwidth}
\includegraphics[angle=0,width=\textwidth]{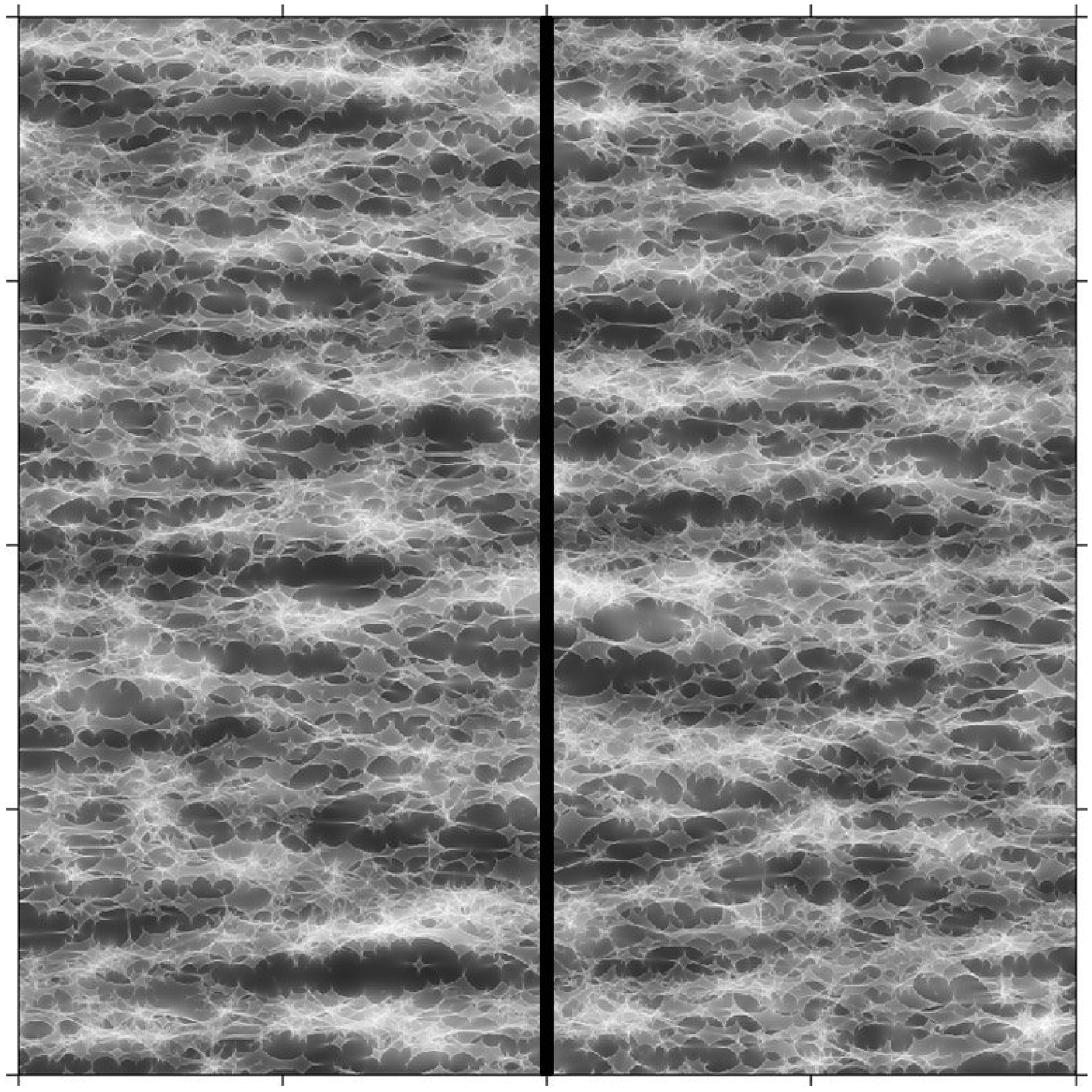}
\end{minipage}
\begin{minipage}[c]{0.5\textwidth}
\includegraphics[angle=0,width=\textwidth]{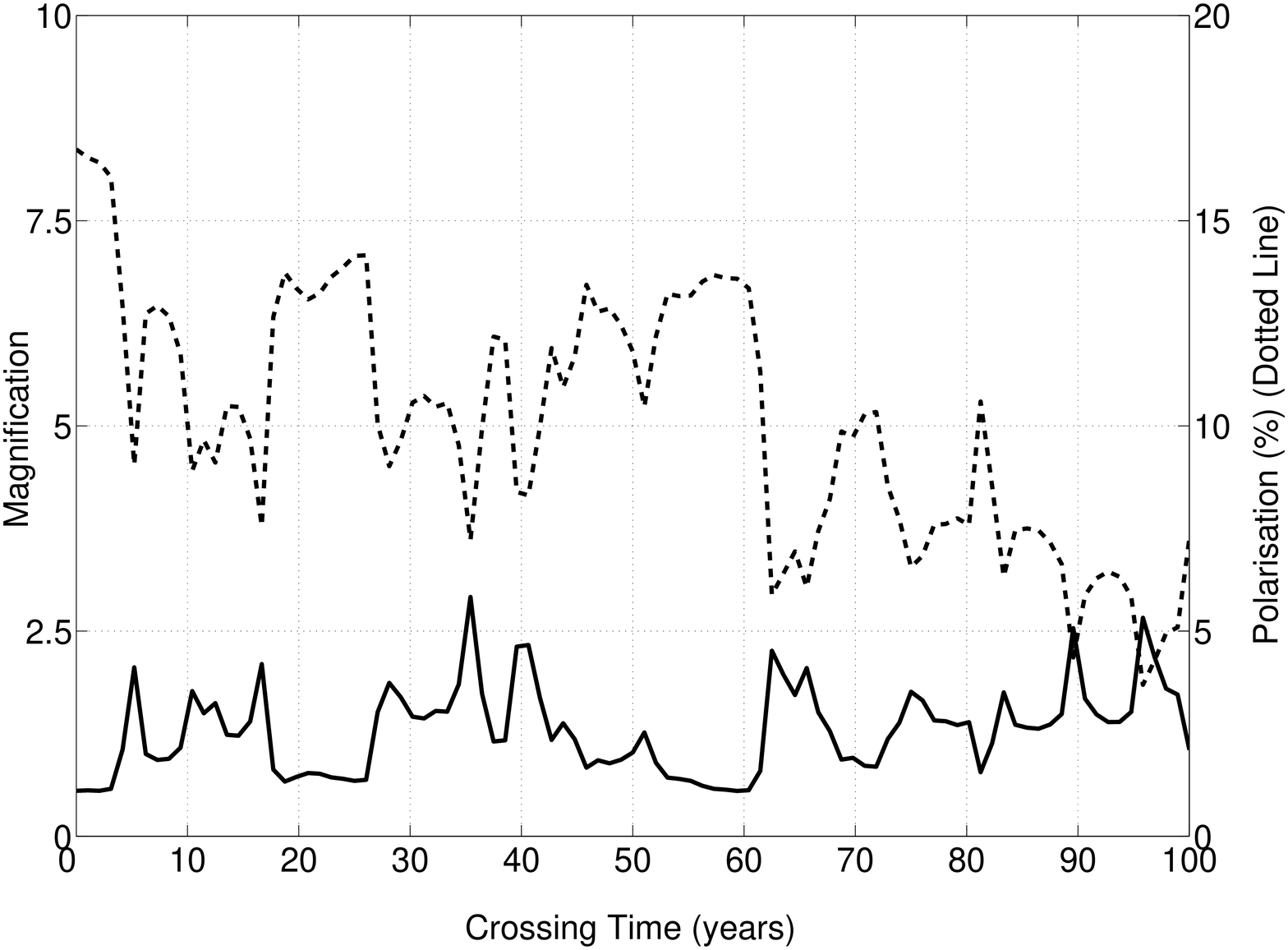}
\end{minipage} \\
\begin{minipage}[c]{0.52\textwidth}
\includegraphics[angle=0,width=\textwidth]{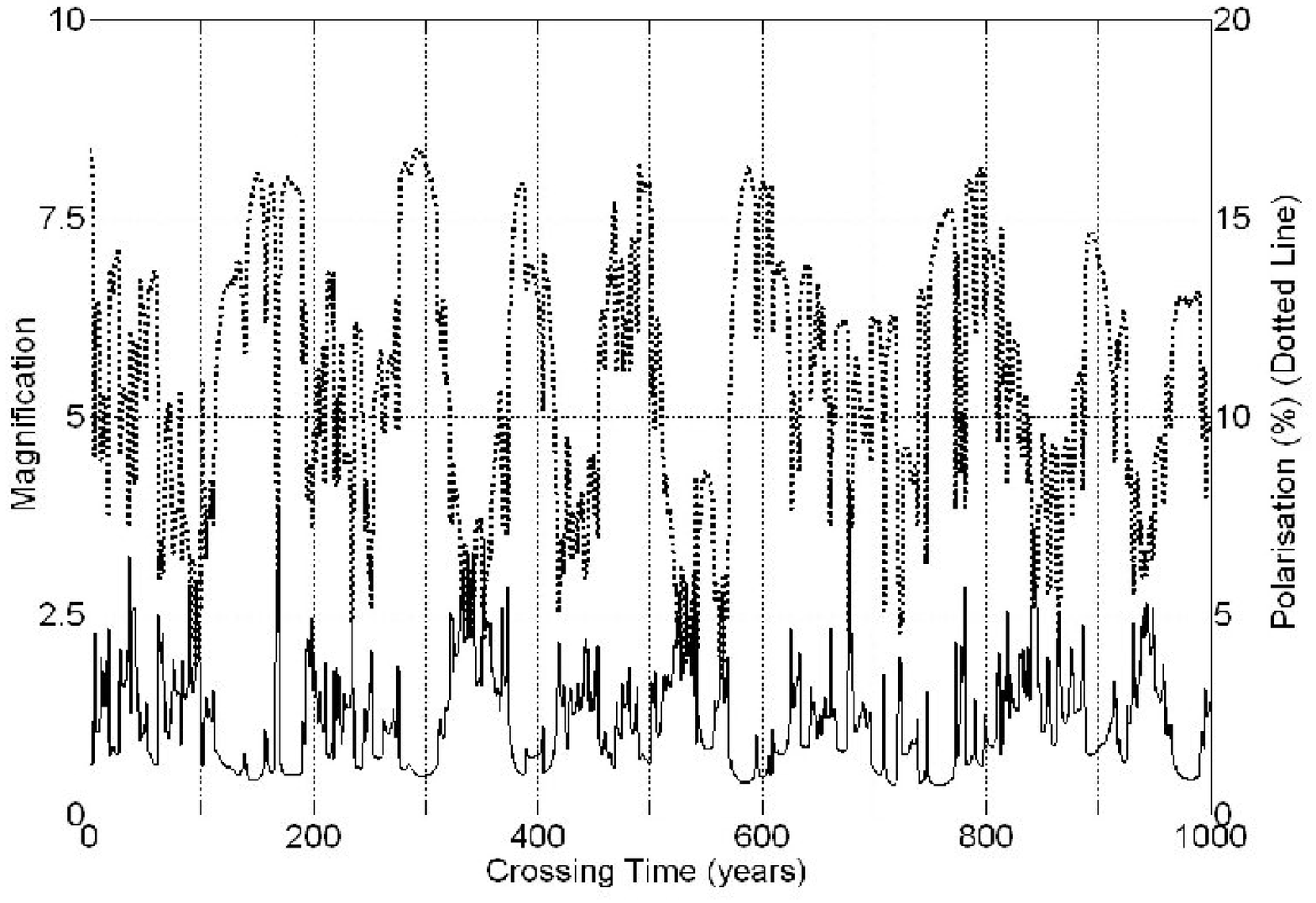}
\end{minipage}
\caption{Model  B. Top Left:  Convolution   of  unabsorbed  source   with  100x100 ER
  magnification  map (scattering  region radius  = 3x$10^{16}$  cm, separation
  distance  = 1.66x$10^{17}$  cm and  $\sigma=\gamma=0.4$). Top Middle: Resultant
  polarisation after application of  Equation \ref{eq6} to two convolutions of
  the   polarised   source   (absorption=91\%). Both have side length 2.56x$10^{18}$ cm.
  Top Right: Zoomed in view of bottom plot, corresponding to first 100 years.
  Bottom:  Light   curves   for
  magnification and polarisation,  as indicated by line from  top to bottom in
  top left and top middle images. Note that the axes are slightly different to
  those in Figure \ref{fig10}.}\label{fig11}
\end{center}
\end{figure*}

Before looking at  any correlation data it is insightful to  look at the light
curves to see just how the  mean magnification and polarisation vary, as shown
in Figures  \ref{fig10} and \ref{fig11}.  These light curves are  generated by
taking  a slice  through the  convolved data,  representing what  an observer
would see as caustics pass over the source.

The time  scales shown  in these figures  have been calculated using Equation
\ref{eq7},  where the  scale  size  of the  source  is $f_{15}$×$10^{15}$  cm,
$z_{l}$ is the  redshift of the lensing galaxy ($z\sim1.55$), and $300v_{300}$ \kms
is the velocity of the microlensing stars  across the  line of sight \citep{lew:1}.
\begin{equation}
\tau\approx\frac{f_{15}}{1+z_{l}}\frac{D_{l}}{D_{s}}\frac{1}{v_{300}}yr\label{eq7}
\end{equation}

However, it is worth noting that the uncertainty in the redshift of the lensing galaxy  results in an
uncertainty on the  Einstein radius in the source  plane and hence the
physical size  of the inferred  absorption and scattering  regions. In
considering  potential lens redshifts  between $z=1$  and $z=2$ (assuming
the cosmology from Section \ref{raytrac}) it can be shown that the  size of
the Einstein  radius varies  roughly linearly
between $+30\%$ and $-30\%$ of the value at $z\sim1.55$. In other words
the time scales shown in Figures \ref{fig10} and \ref{fig11} would vary
by these same percentages.

These results hint at  some interesting correlations between magnification and
polarisation. In  Model A, the  polarisation fluctuations tend to  occur at
caustic crossings and the polarisation seems to oscillate about a mean value. In
Model B  it is  seen  that decreases  in polarisation  tend to occur  at higher
magnification and without any significant oscillation. This can be seen in the top
image of Figure \ref{fig11}, whereby the polarisation during each microlensing event
decreases significantly rather than oscillating about any particular value.
With regard to possible observational testing this
last point is of crucial significance because it indicates that as few as one
major microlensing event may be used to differentiate between the two models. It
is also worth noting that the same number of microlensing magnification events can be statistically
expected for both models within a 20 year period, with this being related to the time
taken for caustic structures to sweep the continuum source. As we will show in the next
section, these trends are confirmed in an analysis of the correlations between
magnification and polarisation for a large sample of microlensing scenarios.

\subsection{Trends in Model A}\label{resA}
\begin{figure*}
\begin{center}
\begin{minipage}[c]{0.98\textwidth}
\includegraphics[angle=-90,width=\textwidth]{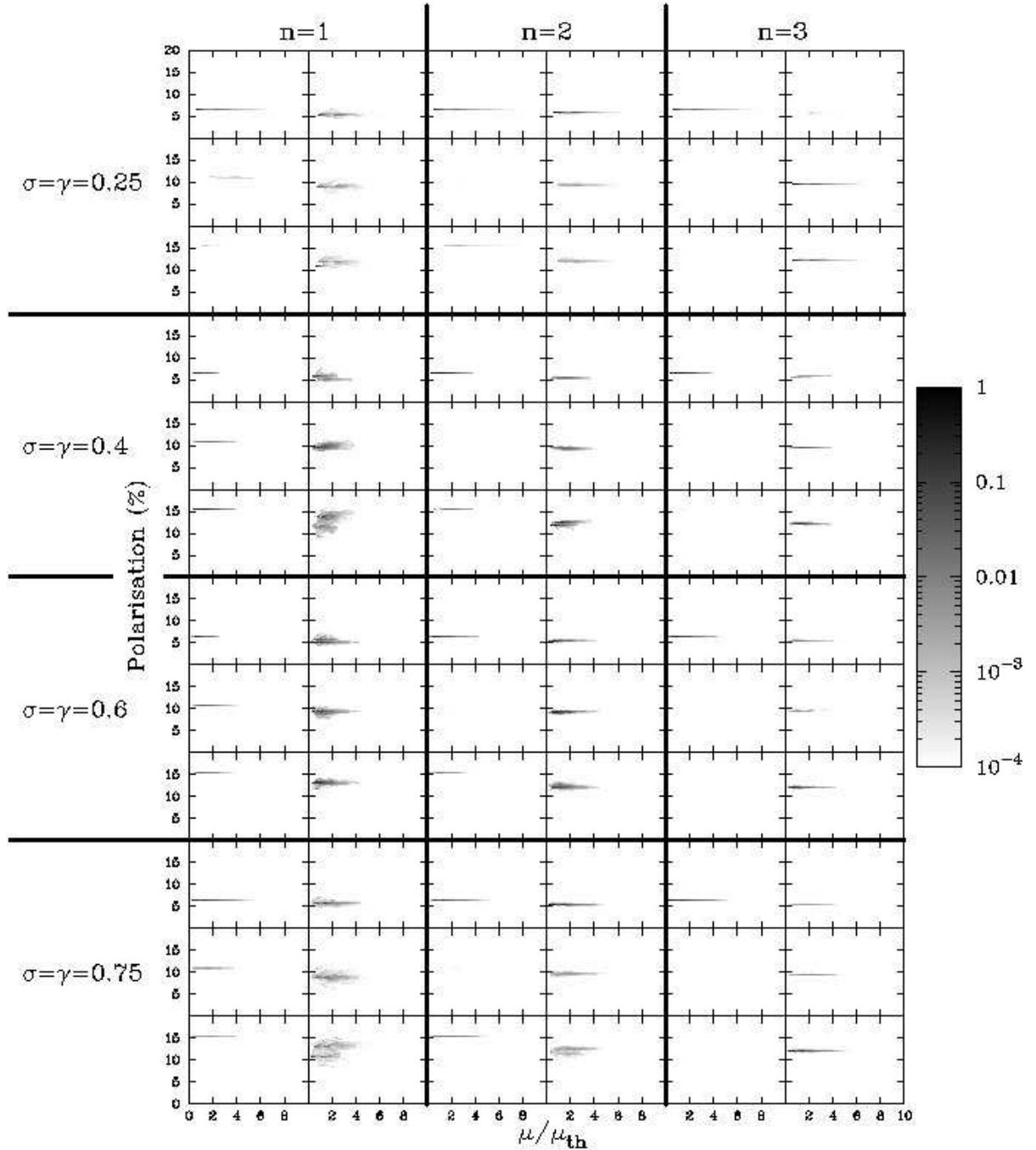}
\end{minipage}
\caption{Correlations  for Model  A.  Note that  the  scale reflects  relative
  probability as denoted  by the grey-scale on the right of  the figure . Each
  group of  6 maps represent mean  absorption at 25\% (top  two), 50\% (middle
  two) and 75\% (bottom two), with an absorption width of 5\% (left three) and
  50\%  (right  three).  For  each   image  the vertical axis shows polarisation
  while the horizontal axis shows magnification (this has been normalised with
  respect to the theoretical mean magnification $\mu_{th}$ (see Equation
  \ref{eq3}).}\label{fig12}
\end{center}
\end{figure*}

The  correlations between  polarisation  and magnification  for the  scenarios
described in  Section \ref{modA} are  shown in Figure \ref{fig12};  each panel
presents  a  grey  scale  map  of relative  probability,  with  the  continuum
magnification on  the  x-axis and  the  degree  of polarisation on the y-axis.
Hence this shows trends between any fluctuations in the two.  Before discussing
the trends seen it should be noted that the apparent decrease in mean polarisation
with increased absorption width (in this  model) is in fact an artifact of the
computational  model.  This  artifact does  not influence  the  overall trends
obtained.

Firstly, it is apparent that  the choice of matter distribution parameters for
the lensing galaxy  (i.e.  $\sigma$ and $\gamma$) do  not significantly affect
the  trends  seen in  these  probability  distributions  (especially with  the
knowledge  that  any  real  data  would  be both  incomplete  and  have  error
associated with it - only the most general of trends would be noticeable). The
same can be  generally said for the  choice of cloud size, although  it can be
seen  that the  range of  polarisations observed  at any  particular  value of
magnification decreases  with decreasing cloud  size. It is also  worth noting
that the  choice of mean absorption does  not induce any trends  into the data
apart from an obvious increase  in mean polarisation, indicating that possible
spectropolarimetric monitoring would not  reveal different trends at different
wavelengths within  the C$_{\textrm{IV}}$ trough (recall  Equation \ref{eq4}).
In addition, by using two sets of data for each correlation map (corresponding
to a  small  and large  mean  for  the magnification  map  used,  see  Section
\ref{conv}), as seen more noticeably for the cases where  n=1 and $\omega$=50\%,
it  can be  seen that  the overall trends developed do not alter with lower or
higher magnification.

All of the scenarios tested seem to support the general trend that an increase
in magnification  of the continuum  flux (i.e. the unabsorbed  continuum) will
not indicate an increase or decrease in mean polarisation, but rather that the
polarisation  will  remain  reasonably  constant throughout  any  microlensing
activity. For  all cases it is  clear that increasing the  width of absorption
will increase the range of polarisations found at any particular magnification
(the largest ranges are found around  a magnification of 1, which of course is
the approximate mean value of  the magnification map).  This then implies that
if  the  source  was  observed  over  several months,  the  largest  range  of
polarisations measured  would have been occurring when  the mean magnification
of the  continuum flux was  around the theoretical  mean.  In other  words,
provided this model represents
the view of  the quasar, then the largest fluctuations  in polarisation would occur
when the continuum flux was at its time-averaged mean.

\subsection{Trends in Model B}
\begin{figure*}
\begin{center}
\begin{minipage}[c]{0.98\textwidth}
\includegraphics[angle=-90,width=\textwidth]{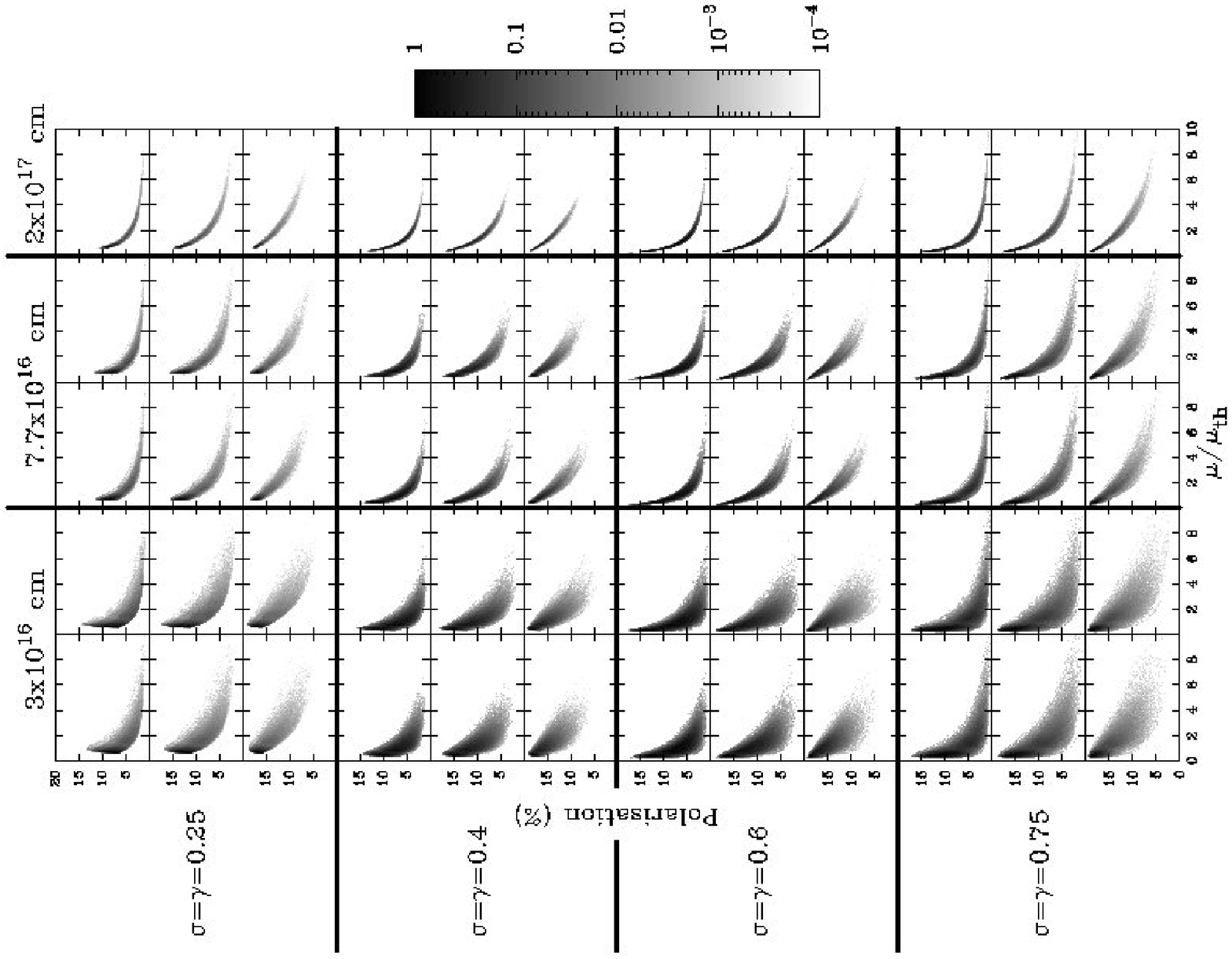}
\end{minipage}
\caption{Correlations for Model  B. The images along each  row represent the 5
  scenarios presented in Figure  \ref{fig9}, namely a scattering region radius
  of  3x$10^{16}$  cm  with   a  separation  distance  of  6.03x$10^{16}$  and
  1.66x$10^{17}$  cm,   7.7x$10^{16}$  cm   with  a  separation   distance  of
  2.00x$10^{17}$ and  5.39x$10^{17}$ cm, and 2x$10^{17}$ cm  with a separation
  distance of 3.70x$10^{17}$  cm. Rows are then arranged  in 3's, representing
  absorption at  77\% (top), 91\% (middle)  and 97\% (bottom).  For each image
  the vertical axis shows polarisation while the horizontal axis shows
  magnification.}\label{fig13}
\end{center}
\end{figure*}

The  correlations between  polarisation  and magnification  for the  scenarios
described  in Section  \ref{modB}  are  shown in  Figure  \ref{fig13}.  It  is
immediately apparent  that the  trends seen in  this model are  very different
from  those seen  in Model  A,  and at  low magnification  (for all  scenarios
tested),   the  polarisation   values  are   much  greater   than   at  higher
magnifications.  Not surprisingly  this is because of the  the geometry of the
source  as  only  the  continuum  source is  strongly  magnified,  whereas  the
scattering region is effectively immune to the influence of the microlenses.

\begin{figure*}
\begin{center}
\begin{minipage}[c]{0.7\textwidth}
\includegraphics[angle=-90,width=\textwidth]{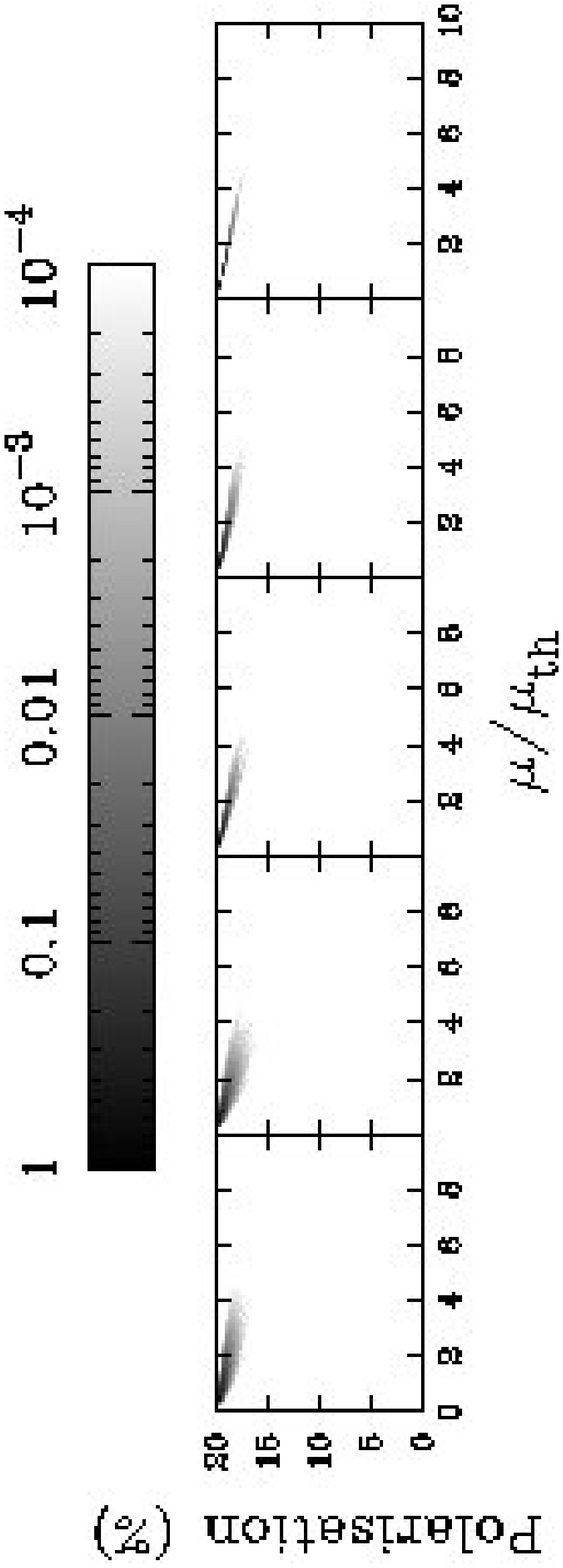}
\end{minipage}
\caption{Correlations  for Model B.  Same as  the $\sigma=\gamma=0.4$  rows of
  Figure \ref{fig13}, but with each image at 99.7\% absorption.}\label{fig14}
\end{center}
\end{figure*}

As with  Model A,  differences between mass  distributions parameters  for the
lensing  galaxy do not  significantly affect  the overall  trends seen  in the
data.   However, unlike  Model  A  it can  be  seen that  the  choice of  mean
absorption does introduce  a specific trend (other than  simply increasing the
mean polarisation); namely that at  low absorption the mean polarisation drops
with increasing magnification while at higher absorption the mean polarisation
tends not  to drop as much. In  fact, at very high  absorption this correlation
begins to look like a horizontal  line. This is illustrated further in Figure
\ref{fig14} (as per a row of Figure \ref{fig13}) for $\sigma=\gamma=0.4$ (this
trend is the  same for the other mass distributions),  where absorption is now
99.7\% (corresponding to a wavelength of $\sim$1530 \AA). This result is important
because it indicates that spectropolarimetric monitoring {\it would} reveal different
trends at different wavelengths within the C$_{\textrm{IV}}$ trough.

Finally, it can  be seen that by increasing the size  of the scattering region
the   relationship  between  magnification   and  polarisation   becomes  more
one-to-one (less broad).  This occurs because, as the  ratio between continuum
and  scattering region  scale sizes  decreases, the  unpolarised  continuum is
allowed to  be magnified in a  more dominant manner thus  reducing the overall
polarisation. It  can also be seen  that variation in  the separation distance
between  continuum  and  scattering  region  centre  does  not  introduce  any
significant trends into  the data. This makes sense, because  as soon as these
two regions  are asymmetrically separated by  at least the width  of a caustic
then the manner in which  they are magnified becomes independent, yielding the
similar results seen here.

\section{Conclusions}
This  study  has   investigated  the  role  that  microlensing   can  play  in
differentiating  between two  popular models  of  how quasar  BAL troughs  are
polarised. Using the  macrolensed and microlensed quasar H1413+1143  as a case
study, two  computational models  of polarisation were  developed in  order to
investigate  how  magnification  and  polarisation variation  detected  by  an
observer could  be used  to differentiate between  them.   The results
showed that  the correlations  between these two  observables would  be easily
discernible between models. Two main differences were identified; for Model A,
during a single microlensing magnification event, the polarisation at any
wavelength within the C$_{\textrm{IV}}$ trough was seen to oscillate about
a mean value, while away from high magnification events the polarisation
was found to be relatively constant. This was found to be in stark contrast with Model B,
whereby the mean polarisation at any particular
wavelength within the trough was found to rise and fall in anti-correlated
fashion with magnification. The second main difference was that if
polarisation and magnification variation were monitored at various
wavelengths within the C$_{\textrm{IV}}$ trough, then significantly different trends would be
seen in this data between wavelengths for Model B but not Model A. Hence
spectropolarimetric monitoring through even a single high magnification event
would provide constraints on the underlying scattering geometry.
Given that caustic crossing times for H1413+1143 would be of order months to years,
an observational campaign monitoring this system on a weekly to monthly basis would
be required to differentiate between the two models. Note however that
spectropolarimetric monitoring is required for high magnification events and an
observational program to obtain high temporal sampling could be triggered based
on simple photometric monitoring of the images.

In  addition, through  the  larger  variations in  polarisation  seen for  all
modelled scenarios  of Model B, this study supports recent data for H1413+1143
\citep{chae:2,char} which also suggests that a scattering region is most likely
responsible  for  the  increased   polarisation  found  in  BAL  troughs. Such
increased variation comes about in this model because the continuum region would
be much more susceptible to microlensing than the larger scattering region. If
this scenario is correct, then this hints that such a scattering region is also
operating in  other  such  quasars. For further work, the  detailed   temporal
properties  of  the expected polarisation  fluctuations need to  be determined
with  the  goal  of  developing  the  most efficient observational strategy to
observe microlensed BAL quasars.

\section*{Acknowledgements}

The anonymous referee is thanked for comments that improved the clarity of this
paper. This research was undertaken as a third year special project at the University
of Sydney.

\end{document}